\documentclass{aa}  

\usepackage{graphicx}
\usepackage{txfonts}

\usepackage{siunitx}

\usepackage[dvipsnames]{xcolor}

\DeclareSIUnit \h {\ensuremath{\mathit{h}}}
\DeclareSIUnit \parsec {pc}
\DeclareSIUnit \msol {\ensuremath{M_{\odot}}}
\DeclareSIUnit \au {AU}
\DeclareSIUnit \year {yr}

\newcommand{\rev}[1]{#1}
\newcommand{\revcom}[1]{}

\newcommand{\myvec}[1]{\vec{#1}}
\newcommand{\mat}[1]{\tens{#1}}

\newcommand{\expect}[1]{\left\langle {#1} \right\rangle}
\newcommand{\expectgal}[1]{\left\langle {#1} \right\rangle_{\mathrm{g}} }
\newcommand{\trace}[1]{\mathrm{tr}\left( {#1} \right)}

\newcommand{\citetprob}[0]{\citet{stucker_2024_prob}}
\newcommand{\citepprob}[0]{\citep{stucker_2024_prob}}

\newcommand{\quotes}[1]{``#1''}

\begin{document}

   \title{Gaussian Lagrangian Galaxy Bias}


   \author{Jens Stücker
          \inst{1}
          \and
          Marcos Pellejero-Ibáñez \inst{2}
          \and
          Rodrigo Voivodic
          \inst{1}
          \and
          Raul E. Angulo \inst{1,3}
          }

   \institute{Donostia International Physics Center (DIPC),
              Paseo Manuel de Lardizabal 4, 20018 Donostia-San Sebastian, Spain\\
              \email{\rev{jens.stuecker@univie.ac.at}}
         \and
             Institute for Astronomy, University of Edinburgh, Royal Observatory, Blackford Hill, Edinburgh, EH9 3HJ , UK\\
             \email{mpelleje@ed.ac.uk}
         \and
             IKERBASQUE, Basque Foundation for Science, E-48013, Bilbao, Spain
             }

   \date{March 2024}

 
  \abstract
   {Understanding \emph{galaxy bias} -- that is the statistical relation between matter and galaxies -- is of key importance for extracting cosmological information from galaxy surveys. While the \quotes{bias function} $f$ -- that is the probability of forming galaxies in a region with a given density field -- is usually approximated through a parametric expansion, we show here, that it can also be measured directly from simulations in a non-parametric way. Our measurements show that the Lagrangian bias function is very close to a Gaussian for halo selections of any mass. Therefore, we newly introduce a Gaussian bias model with several intriguing properties: (1) It predicts only strictly positive probabilities $f > 0$ (unlike expansion models), (2) It has a simple analytic renormalized form and (3) It behaves gracefully in many scenarios where the classical expansion converges poorly. We show that the Gaussian bias model describes the galaxy environment distribution $p(\delta | \mathrm{g})$, the scale dependent bias function $f$ and the renormalized bias function $F$ of haloes and galaxies generally equally well or significantly better than a second order expansion with the same number of parameters. We suggest that a Gaussian bias approach may enhance the range of validity of bias schemes where the canonical expansion converges poorly and further, that it may make new applications possible, since it guarantees the positivity of predicted galaxy densities.
   }


   \maketitle
%

\section{Introduction}

The spatial distribution of galaxies is one of the most promising probes of the cosmology of our Universe. While their clustering through gravity can be modelled accurately through perturbation theory \citep[see][for a review]{Bernardeau_2002} and N-body simulations \citep[see][for a review]{AnguloHahn2022}, their formation and morphological evolution is a highly complicated process that is so far difficult to predict reliably. Accounting for this appropriately is one of the central challenges for optimally interpreting current and future large scale surveys, like the Dark Energy Instrument (DESI, \citealt{DESI}), or Euclid (\citealt{Euclid}).

Despite the complexity of galaxy formation, the clustering of galaxies appears rather simple on sufficiently large scales. For example, on linear scales, the two point correlation function of galaxies can be well described by a linear bias factor times the correlation function of matter \citep{kaiser_1984}. Such simplicity arises because the large scale clustering is mostly driven (beyond the gravitational movement) by the average response of large ensembles of galaxies to small perturbations in the linear density field. This response can be modelled through a simple bias relation \citep[see][for a review]{Desjacques_2018}.

In bias expansion approaches, the bias function is expanded as a Taylor series in terms of small perturbations to the linear fields. This approach offers great flexibility, allowing a single model to describe biased tracers with vastly different properties down to $k\approx 0.2h/$Mpc \citep{Baumann_2012,Baldauf_2016,Vlah_2016}.
Bias expansion approaches have been applied successfully to extract robust cosmological constraints from surveys \citep{Ivanov_2020,d'Amico_2020,Colas_2020,Nishimichi_2020,Chen_2020,Philcox_2022}.

However, traditional bias methods also exhibit some noteworthy inconsistencies. Bias expansion models generically predict negative galaxy densities for some values of the underlying matter densities \citep[e.g.][]{Wu_2022}. While the amplitude of this problem can in principle be controlled by limiting the investigation to very large scales with small variance, it is almost inevitable that some high-$\sigma$-outlier regions of space are predicted to be negative. This fact is commonly ignored since the summary statistic of interest -- like e.g. the power spectrum or the field level error spectrum \citep{Schmittfull_2019} -- may not be significantly affected by it. However, the predicted negative densities automatically exclude some possible applications of bias models. For example, it prevents describing the locations of galaxies through their joint probability distribution, or it makes it impossible to consistently sample discrete tracers from the predicted galaxy density field.

Further, it is not always clear, how well the bias expansion converges. For example, the coefficients of higher order terms may actually grow so that not every perturbative series is well convergent \citepprob{}. This can severely limit the ability to describe highly biased objects even at quite large scales.


It is the goal of this article to investigate the properties of the bias function in a \emph{non-parametric} way and to propose a solution to the mentioned shortcomings through a Gaussian Lagrangian bias model.

Non-perturabtive and strictly positive formulations of the bias function have already been considered in previous studies \citep{Szalay_1988, Matsubara_1995, Sigad_2000, Matsubara_2011, Neyrinck_2014, Friedrich_2022}. Such approaches have been particularly popular in reconstruction methods where a likelihood for the joint distribution of all observed galaxies needs to be modelled \citep{Ata_2015, jasche_2019, Hernandez_2021}. For example, the \textsc{BORG} and \textsc{COSMIC BIRTH} algorithms \citep{jasche_2019,Kitaura_2021} uses a powerlaw bias function with exponential truncation in low density regions developed by \citet{Neyrinck_2014}. These bias models are considerably more complicated than the one that we propose here, because they are formulated in Eulerian space where the matter distribution is already quite complex.

Analytical models of structure formation give naturally rise to strictly positive formulations of the Lagrangian bias function: In excursion set models, bias parameters can be estimated through the probability of crossing a small scale barrier given a large scale perturbation \citep{bond_1991, Mo_White_1996, musso_2012}. In peak theory, the response of the number density of peaks can be investigated as a function of a larger scale density perturbation \citep{bardeen_1986}. In both cases are the resulting bias relations closely related to the Gaussian statistics of the linear density field and therefore close to a Gaussian. From an analytical perspective it seems natural to consider parametric approaches that easily recover a Gaussian form for the bias function. \revcom{This paragraph is new.}

The bias relation has already been investigated by non-parametric methods in previous studies. On the one hand, there exist approaches that smooth both the matter density field and the galaxy density field (both in Eulerian space) and study the relation through the \quotes{scatter-plot method}  \citep{manera_2011, uhlemann_2018, Desjacques_2018,Balaguera_2019,Pellejero_2020,Kitaura_2022,Friedrich_2022}. This approach is relatively complex, because the bias relation depends on both smoothing scales and further stochasticity needs to be described simultaneously to the deterministic aspects of the bias relation. On the other hand, recently another non-parametric approach has been proposed by \citet{Wu_2022} where a high dimensional Lagrangian bias function $f$ is defined on a grid and is fitted to reproduce the Eulerian galaxy density field \citep[see also][]{Wu_2023}. The authors show that this approach gives a consistent bias model that follows the physical constraints ($f > 0$) and accurately describes the galaxy field. However, this comes at the cost of a very high dimensional parameter space and a very complex optimization problem.

Here, we present a new method for measuring the Lagrangian bias function $f$. We define $f$  as the excess probability of forming a galaxy in an infinitesimal Lagrangian volume element in dependence on the properties of the linear density field. The measurement of $f$ is done through the distribution of linear densities at the Lagrangian locations of galaxies (or haloes) $p(\delta | \mathrm{g})$ -- that is the \quotes{galaxy environment distribution}. Our approach is considerably simpler than previous approaches and it only requires smoothing the linear density field, but not the galaxy field. Further, we show that while the bias relation $f$ is dependent on the smoothing scale that it is measured at, it is possible to define and measure a renormalized bias function $F$, which captures those aspects of $f$ that are independent of the measurement scale.

Further, we newly introduce a Lagrangian Gaussian bias model. 
We show that the Gaussian bias model has intriguing properties that distinguish it from a traditional bias expansion, most noteworthy that it predicts only strictly positive probabilities $f > 0$. We show, that the Gaussian bias model qualifies as a valid bias model, since it can be written in a renormalized form that is mutually consistent across different smoothing scales. Further, we show that in the multivariate case (e.g. considering additionally the tidal field or the Laplacian of the density field) a straightforward generalization is given by a multivariate Gaussian.

We measure the bias function $f$, the galaxy environment distribution $p(\delta | \mathrm{g})$ and the renormalized bias function $F$ for a large variety of cases -- including haloes of different mass selections and mock galaxy catalogues. We find that the bias function of haloes is extremely close to a Gaussian and almost perfectly described by the Gaussian bias model -- both in the monovariate and the multivariate case. We show that a Gaussian can give a much improved description compared to an expansion bias model with the same number of parameters -- especially for highly biased objects that are very challenging to describe with the traditional approach. 

The considerations in this article are limited to Lagrangian space, since only in this case  the distribution of matter is known exactly and given by a multivariate Gaussian. They combine therefore optimally with hybrid bias approaches where bias functions are set up in Lagrangian space and advected to Eulerian space through the displacement field from N-body simulations \citep{Modi_2020,Kokron_2021,Zennaro_2023,Pellejero_2022,Pellejero_2023,DeRose_2023}. 

We publish this article jointly with a companion paper, \citetprob{} in which we investigate bias parameters through the moments and cumulants of the galaxy environment distribution. One of the main results of that article is that cumulant biases beyond order three are close to zero which additionally motivates the consideration of the Gaussian bias function which we present here.

The article is structured as follows: In Section \ref{sec:theory}, we introduce for the mono-variate density-only case the necessary definitions and theory to understand and measure bias at a functional level. Further, we introduce the Gaussian bias model and show that it can be written in a self-consistent renormalized form. In Section \ref{sec:measure_monovariate} we show measurements of monovariate bias functions and we compare them to the Gaussian and a second order expansion model. In Section \ref{sec:multivar_theory}, we extend the theory to multiple variables and in Section \ref{sec:measure_multivariate} we show measurements of the multivariate bias function. In Section \ref{sec:measureF} we show how bias functions across multiple scales can be described by a single multivariate model and show how this can be summarized through the scale-independence of the renormalized bias function. In Section \ref{sec:whygaussian} we give a physical interpretation of our measurements and speculate about the reason why the bias function appears to be Gaussian in so many scenarios. Finally, in Section \ref{sec:conclusions} we summarize our discoveries and suggest possible applications.


\section{Theory} \label{sec:theory}
\begin{figure*}
    \centering
    \includegraphics[width=\textwidth]{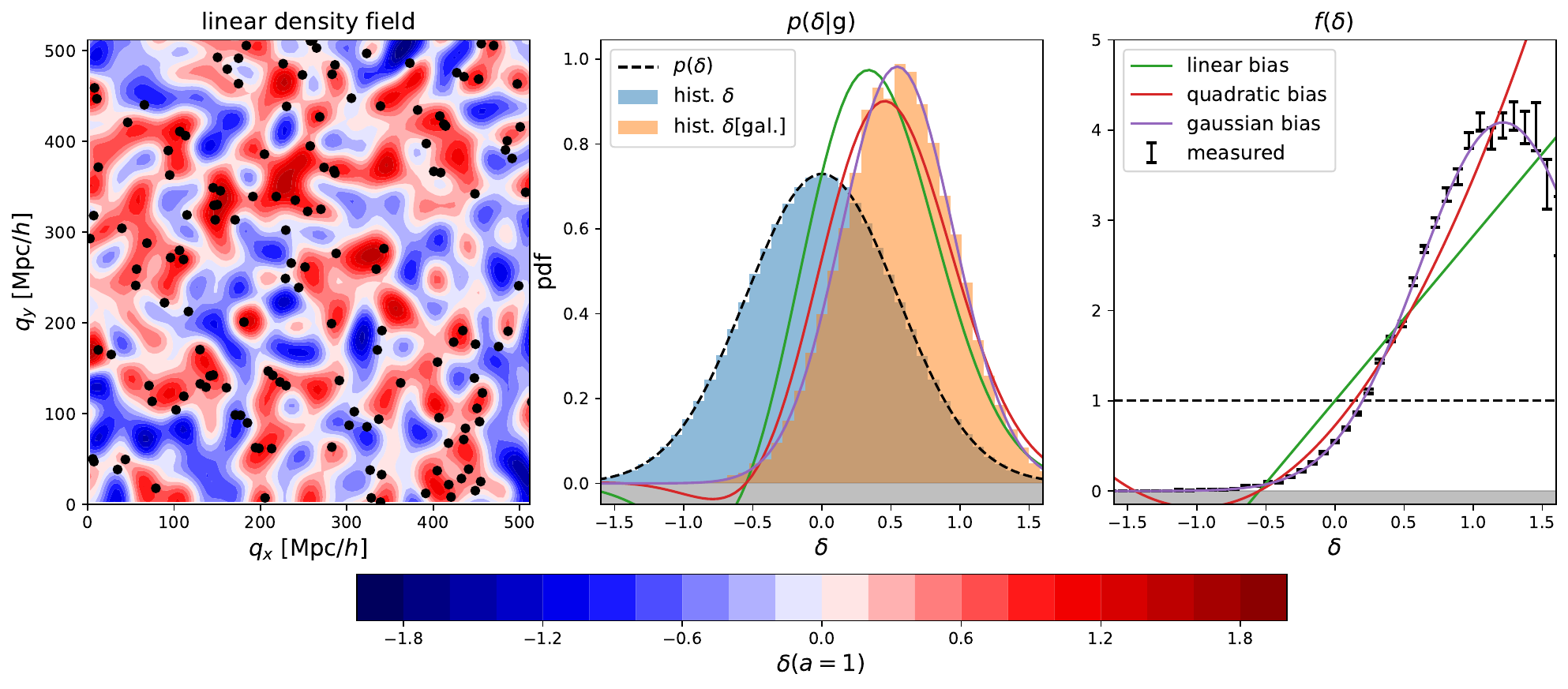}
    \caption{Illustration of the steps necessary for measuring the bias function $f$. Left: the smoothed linear density field and the initial (Lagrangian) locations of a set of tracers -- here the most bound particles of haloes with $M_{200\rm{b}} \sim \SI{2e14}{\per \h \msol}$. Center: Distribution of the linear densities at the tracer locations (orange histogram) which is biased relative to the distribution at random locations (blue histogram). Right: The bias function, which is inferred by dividing the orange histogram by the background distribution. The solid lines show different bias models out of which the Gaussian bias model describes the data clearly the best.}
    \label{fig:bias_function}
\end{figure*}

In this Section, we will show how to measure the (scale dependent) bias function in a non-parametric way through the galaxy environment distribution. Further, we show how to relate the scale dependent bias function to its ``renormalized'' large scale limit and how to define parametric forms with scale independent bias parameters. We newly introduce the Gaussian bias model as a compelling parametric form of the bias function. For simplicity, the considerations in this section limit only to Lagrangian density bias and we will consider the more general multivariate case later in Section \ref{sec:multivar_theory}. Many of the considerations here are brief summaries of the more elaborate derivations presented in \citetprob{}.

\subsection{Measuring the bias function}

\noindent Arguably, the most insightful way to understand galaxy bias is by measuring the bias function directly. Here, we propose a new non-parametric method for measuring the Lagrangian bias function $f$ from cosmological simulations.

We consider an infinitesimally small Lagrangian volume element which nothing is known about except the linear density contrast $\delta$ smoothed at some scale (and possibly other features of the linear field like the Laplacian $L$ or the tidal field). Neglecting primordial non-Gaussianity, the density contrast follows a Gaussian distribution 
\begin{align}
    p(\delta) &= \frac{1}{\sqrt{2 \pi} \sigma} \exp \left( - \frac{\delta^2}{2 \sigma^2} \right) \,\,. \label{eqn:p_of_delta}
\end{align}
For simplicity, we will assume throughout this article that the smoothed density contrast is defined with a sharp Fourier-space filter. However, we discuss in the Appendix of \citetprob{} how other filters can easily be incorporated by taking additionally into account the effect of a slightly different correlation between large and small scales. 

 We call the average probability that a galaxy forms in such a volume element $p(\mathrm{g})$ and we call the conditional probability, given the knowledge of the linear density contrast, $p(\mathrm{g} | \delta)$. The excess probability
\begin{align}
    f(\delta) := \frac{p(\mathrm{g} | \delta)}{p(\mathrm{g})}
\end{align}
is parameterized through a function $f(\delta)$ which we will refer to as the \quotes{scale dependent bias function} or just the \quotes{bias function} throughout this article. The bias function depends in a predictable manner on the variance of $\delta$ at the considered scale, as we will see in Section \ref{sec:theory_Frenorm}.

Further, we define the \quotes{galaxy environment distribution} as the distribution of (smoothed) linear densities at the Lagrangian locations of galaxies $p(\delta | g)$. This distribution is related to the bias function through Bayes' theorem:
\begin{align}
    f(\delta) = \frac{p(\rm{g}|\delta)}{p(\rm{g})} = \frac{p(\rm{g} \cap \delta)}{p(\rm{g}) p(\delta)}  = \frac{p(\delta | \rm{g})}{p(\delta)} \label{eqn:f_versus_pg} \,\,.
\end{align}
That is, the bias function is the ratio between the galaxy environment distribution $p(\delta | \rm{g})$ and the background distribution $p(\delta)$. We can use this fact to measure $f$ in the following steps:


In a simulation, we can trace galaxies or haloes back to Lagrangian space $\myvec{q}_i$. For example, we consider the set of tracers that is given by the locations of the most bound particles of haloes with a mass close to $M_{200\rm{b}} \sim \SI{2e14}{\per \h \msol}$. The Lagrangian positions $\myvec{q}_i$ of these tracers can be inferred from the ids of the most bound particles. We show the Lagrangian positions of such tracers in the left panel of Figure \ref{fig:bias_function}. Further, we consider the smoothed linear density field\footnote{with a damping scale $k_{\rm{d}} = \SI{0.15}{\h \per \mega \parsec}$ as explained in Section \ref{sec:measure_monovariate}} $\delta(\myvec{q})$ which is shown as coloured contours in the left panel of Figure \ref{fig:bias_function}. 

The galaxy environment distribution $p(\delta | g)$ is now simply given by the distribution of the linear density contrast at the locations of these tracers $\delta_i = \delta(\myvec{q}_i)$. 
We show this as an orange histogram
in the central panel of Figure \ref{fig:bias_function}. If the tracers were uniformly distributed in Lagrangian space, they would follow a Gaussian distribution as in equation \eqref{eqn:p_of_delta} -- indicated as a black dashed line and as a blue histogram in the central panel of Figure \ref{fig:bias_function}. However, the distribution of galaxies is notably biased with respect to the distribution of matter.

As shown in equation \eqref{eqn:f_versus_pg}, we can divide the galaxy environment distribution by the background distribution to infer the bias function.  We show this as black data points in the right panel of Figure \ref{fig:bias_function} (with jackknife error bars as will be explained in Section \ref{sec:lag_fields}). We show the bias function in comparison to a linear, a quadratic and a Gaussian approximation. Strikingly, the bias function for the selected set of haloes is very close to a Gaussian. We will show in Section \ref{sec:measure_monovariate} that this is the case for many different scenarios. This motivates to approximate the bias function through a Gaussian bias model.

In the remainder of this section, we will show how to write such a Gaussian bias model in renormalized form with scale independent bias parameters, by relating the scale dependent bias function $f$ to its renormalized large scale limit.

We note that our method for inferring the bias function through a histogram is considerably simpler than previously presented non-parametric approaches \citep{Wu_2022, Wu_2023}. For example, \citet{Wu_2022}, have inferred the bias function by fitting a large number of function values as free parameters to optimally recover the Eulerian galaxy density. In this article, we will mostly use our new method to investigate what are good parametric approximations to the bias function. However, this method can also have important other applications. For example, one could measure the non-parametric bias function $f$ in a small hydrodynamical simulation and use it in larger N-body simulations to create mock catalogues of galaxies that are biased in a similar way.

\subsection{The renormalized bias function $F$} \label{sec:theory_Frenorm}
The bias function $f$ is strongly dependent on the smoothing scale at which it is inferred. However, it is possible to define a scale independent bias function $F$ which corresponds to the large scale limit of $f$. 

The considerations here are based on the idea of the peak-background split (PBS) which states that ``a long-wavelength density perturbation acts like a local modification of the background
density for the purposes of the formation of halos and galaxies' \citep{kaiser_1984, bardeen_1986, Desjacques_2018}. Bias parameters describe the response of the galaxy number to such long-wavelength perturbations.

An exact implementation of the PBS is given by the separate universe approach. In this approach, one considers some universe in which a measurable number of galaxies $n_{g,0}$ forms. If one were to increase the background density of the universe by a relative amount $\delta_0$ (e.g. in a separate universe simulation, \citealt{li_2014, wagner_2015}), then in the new universe a different number of galaxies $n_{g}(\delta_0)$ forms. We call their ratio
\begin{align}
    F(\delta_0) &= \frac{n_g(\delta_0)}{n_{g,0}} = 1 + b_1 \delta_0 + \frac{1}{2} b_2 \delta_0^2 + ...
\end{align}
the \quotes{large scale limit of the bias function} or the \quotes{renormalized bias function}. This function can directly be measured with separate universe simulations \citep[e.g.][]{lazeyras_2016} and we will refer to the coefficients of the indicated expansion as the \quotes{canonical bias parameters} or just \quotes{the bias parameters}:
\begin{align}
    b_n &=  \left. \frac{\partial^{n} F(\delta_0)}{\partial \delta_0^{n}} \right|_{\delta_0=0} \,\,.
    \label{eqn:pbs_bias}
\end{align}
These bias parameters physically describe the response of the galaxy number density to small perturbations at infinitely large scales. However, in practice, perturbations do not originate from infinitely large scales, but from finite large scales, so that it is necessary to relate $F$ to the scale dependent bias function $f$ that is observable at finite scales.

Since densities at different scales add up linearly, a separate-universe style modification of the large scale density contrast from 0 to $\delta_0$ will immediately translate to a modification of the linear density in every volume element $\delta \rightarrow \delta + \delta_0$. Therefore $F$ and $f$ should be related through
\begin{align}
    F(\delta_0) &=  \langle f(\delta + \delta_0) \rangle \label{eqn:ngversusbias}
\end{align}
where the angled brackets indicate an expectation value taken over the Lagrangian volume \citep[see also][]{Desjacques_2018} so that
\begin{align}
    F(\delta_0) &= \int_{-\infty}^\infty  p(\delta) f(\delta + \delta_0) \mathrm{d} \delta \,\,. \label{eqn:Frenorm}
\end{align}
The relation indicates that, in a separate universe experiment, the number of galaxies should change according to the average change in probability of forming galaxies when changing the linear density contrast everywhere in space. Importantly, this relation defines how to renormalize bias at the functional level, which we will use a lot throughout this article. Note that it corresponds to a convolution with the Gaussian background distribution. Therefore, parametric models which maintain (the same) parameteric form after convolution with a Gaussian qualify as particularly convenient bias models. We will see examples of this in the next two subsections. 

\subsection{Expansion bias}
At second order (in density only) the canonical bias expansion of the large scale bias function $F$ reads
\begin{align}
    F_{\mathrm{quad}}(\delta_0) &= 1 + b_1 \delta_0 + \frac{1}{2} b_2 \delta_0^2 \,\,. \label{eqn:Fquad}
\end{align}
To find a form for the scale dependent bias function $f$, we make a quadratic Ansatz
\begin{align}
    f_{\mathrm{quad}}(\delta) &= c_0 + c_1 \delta + \frac{1}{2} c_2 \delta^2
\end{align}
and we apply equation \eqref{eqn:ngversusbias} to obtain
\begin{align}
    F_{\mathrm{quad}}(\delta_0) &= \expect{c_0 + c_1 (\delta + \delta_0) + \frac{1}{2} c_2 (\delta^2  + 2 \delta \delta_0 + \delta_0^2)} \nonumber \\
                &= c_0 + c_1 \delta_0 + \frac{1}{2} c_2 \sigma^2 + \frac{1}{2} c_2 \delta_0^2
\end{align}
where we have used that for the Gaussian background distribution $\expect{\delta} = 0$ and $\expect{\delta^2} = \sigma^2$. By identifying coefficients with equation \eqref{eqn:Fquad}, we find $c_1 = b_1$, $c_2 = b_2$ and $c_0 = 1 - \frac{1}{2} b_2 \delta_0^2$ leading to
\begin{align}
    f_{\mathrm{quad}}(\delta) &= 1 + b_1 \delta + \frac{1}{2} b_2 (\delta^2 - \sigma^2)
\end{align}
as the renormalized form of the quadratic bias function. Note that polynomials of any degree maintain their degree after convolution with a Gaussian which makes it possible to renormalize any polynomial bias model in a simple manner. Further, it is worth emphasizing that the renormalization procedure is quite simple here, because $\delta$ is the linear density field in Lagrangian space. In Eulerian bias schemes the renormalization procedure can be considerably more complex \citep[e.g.][]{assassi_2015}.

\subsection{Gaussian bias} \label{sec:gaussianbias}
In \citetprob{} we have introduced \quotes{cumulant bias parameters} as an alternative way of phrasing the bias expansion. These are defined as
\begin{align}
    \beta_n &= \left.  \frac{\partial^n}{\partial \delta_0^n}  \log \left( F(\delta_0) \right) \right|_{\delta_0=0} \label{eqn:cum_pbs_bias}
\end{align}
and they are related to canonical bias parameters in the same way that cumulants are related to moments, e.g.
\begin{align}
    \beta_{1} &= b_{1} \label{eqn:beta1_of_b} \\
    \beta_{2} &= b_{2} - b_{1}^{2} \\
    \beta_{3} &=  b_{3} - 3 b_{1} b_{2}  + 2 b_{1}^{3} \,\,. \label{eqn:beta3_of_b} 
\end{align}
In \citetprob{} we have found that cumulant biases are very close to zero at orders $n \geq 3$. This motivates to consider a cumulant bias expansion that is truncated at second order
\begin{align}
    \log F_{\mathrm{gaus}} &= \beta_1 \delta_0 + \frac{1}{2} \beta_2 \delta_0^2 \label{eqn:Fgaus}
\end{align}
as a particularly interesting case. Under the constraint $\beta_2 < 0$, which is generally fulfilled as shown in \citetprob, $F_{\mathrm{gaus}}$ is a Gaussian function that is normalized to $F_{\mathrm{gaus}}(0) = 1$, that has its maximum at $- \beta_1 / \beta_2$ and a width of $1/\sqrt{-\beta_2}$. Again, we can find an explicit form for the scale dependent bias function $f_{\mathrm{gaus}}$ through an Ansatz
\begin{align}
    f_{\mathrm{gaus}}(\delta) &= N_b \exp \left(- \frac{(\delta - \mu_{b})^2}{2 \sigma_b^2}  \right) \\
    F_{\mathrm{gaus}} &= \int_{-\infty}^{\infty} f_{\mathrm{gaus}}(\delta + \delta_0) \frac{1}{\sqrt{2 \pi} \sigma} \exp \left( -\frac{\delta}{2 \sigma^2} \right) \mathrm{d} \delta\\
      &= \frac{N_b}{\sqrt{1 + \sigma^{2} / \sigma_b^2}} \exp \left(-\frac{ \left(\delta_0 - \mu_b\right)^2}{2 (\sigma^2 + \sigma_b^2)}\right) \,\,.
\end{align}
By taking the logarithm and identifying coefficients with equation \eqref{eqn:Fgaus}, we find
\begin{align}
N_b   &= \frac{\exp \left(- \frac{\beta_{1}^{2}}{2 \beta_{2}}\right)}{\sqrt{\beta_{2} \sigma^{2} + 1}} \\
\mu_{b}   &=- \frac{\beta_{1}}{\beta_{2}} \\
\sigma_b^2   &= - \frac{1}{\beta_{2}} - \sigma^{2}
\end{align}
which leads to
\begin{align}
   f_{\mathrm{gaus}} &= \frac{\exp \left(- \frac{\beta_{1}^{2}}{2 \beta_{2}}\right)}{\sqrt{1 + \beta_{2} \sigma^{2}}} \exp \left(\frac{\beta_{2} \left(\frac{\beta_{1}}{\beta_{2}} + \delta\right)^{2}}{2 (1 + \beta_{2} \sigma^{2})} \right)
\end{align}
as the Gaussian density bias model in renormalized form -- our first important theoretical result.

First of all, we note a few consistency properties of this Gaussian bias model. The large scale limit of $f_{\mathrm{gaus}}$ yields $F_{\mathrm{gaus}}$:
\begin{align}
    \lim_{\sigma \rightarrow 0} f_{\mathrm{gaus}} &= \exp \left( \beta_1 \delta + \frac{\beta_2 \delta^2}{2} \right) = F_{\mathrm{gaus}}(\delta) \,\,.
\end{align}
So, under the Gaussian bias assumption, both the bias function that can be deduced from finite scales and the limiting function at infinite scales are Gaussians and mutually consistent with each other. This is an important result, since it ensures that a bias model that is set up to be Gaussian at some small scale, will also appear Gaussian at any larger scale. 

Further, we note the Taylor expansion of $F$ around $\delta_0 = 0$ is consistent with the canonical bias expansion at second order and many of the results that are valid for second order expansion models translate directly to the Gaussian model. However, the Gaussian makes a more graceful assumption about unmodelled higher order terms which ensures improved behavior outside of the $|\delta_0| \sim 0$ regime. For example, the Gaussian ensures $f > 0$ for any amplitudes of density perturbations. Since we should only observe positive probabilities and positive galaxy number densities in the real universe, this is a desirable property. Ensuring it may allow additional applications for bias models, as we will discuss in Section \ref{sec:conclusions}.

Regarding the parameters of the Gaussian, we note that the location of the maximum $\mu_b$ is independent of $\sigma$ and gets larger for larger $\beta_1$ (since $\beta_2 < 0$). The effective width of the bias function $\sigma_b$ decreases as $\sigma$ increases. This is so, since a larger fraction of the information, that is necessary to decide whether a volume element collapses into a halo, is resolved. 

Noteworthy, this may lead to undefined behavior beyond $\sigma_{\mathrm{max}} = 1 / \sqrt{-\beta_2}$. If a model is extrapolated with fixed parameters to the corresponding scale, the bias model becomes a Dirac delta function and galaxy formation becomes formally deterministic. Latest at that scale the PBS assumption has to break down, since density perturbations from smaller scales have to impact galaxy formation in a different way (e.g. they become irrelevant). As we explain in \citetprob, this behavior is not unique to the Gaussian bias model, but it must happen for any model that is (physically correctly) restricted to positive probability densities: The PBS predicts that the galaxy environment distribution has zero variance at the corresponding scale and negative variance beyond that scale. Expansion biases simply hide this problematic fact by allowing negative probability densities and a negative variance. When considering additional variables (like the Laplacian) the corresponding scale can be shifted to smaller scales, but generally there has always to be a scale where the PBS breaks down, because all information has been accounted for. We expect that fitting bias models beyond this point will always lead to scale dependent bias parameters, that ensure e.g. $0 > \beta_2 > - 1 / \sigma^2$ for the pure density bias. We discuss this in more detail in \citetprob.

We summarize that a Gaussian bias model is a well defined bias model that has a simple renormalized form and therefore makes consistent predictions across different damping scales. In the limit of large scales and small values of $\delta$, it gets arbitrarily close to a quadratic bias model, but it ensures additionally $f > 0$ outside the perturbative range. 

Throughout this article we will always compare the performance of a Gaussian bias model to a quadratic expansion bias, since these two models use both two free parameters and have the same limiting behaviour on large scales.

\subsection{Bias parameters}
To compare non-parmetric measurements of the bias function with the bias models, we have to infer the bias parameters $b_1$ and $b_2$ that are used to parameterize them. As shown in \citetprob, these can easily be found through the moments of the galaxy environment distribution
\begin{align}
    b_n 
        &= (-1)^n \int_{-\infty}^\infty  \frac{p^{(n)}(\delta)}{p(\delta)} p(\delta |g) \mathrm{d} \delta   \nonumber \\
        &= (-1)^n \expectgal{ \frac{p^{(n)} (\delta)}{p(\delta)} } \label{eqn:bn_from_pderiv}
\end{align}
where \quotes{$(n)$} in the superscript indicates an $n$-th derivative with respect to $\delta$ and the expectation value with a \quotes{g} in the subscript denotes that this expecation is taken over the galaxy environment distribution. Therefore, this estimator only has to be evaluated at the location of galaxies. Inserting the Gaussian distribution, we find
\begin{align}
    b_n &= \expectgal{ \frac{H_n \left( \delta / \sigma \right)}{\sigma^n} } \label{eqn:bn_o0} \\
    b_1 &= \expectgal{ \frac{\delta}{\sigma^2} } \label{eqn:b1_o0} \\
    b_2 &= \expectgal{ \frac{\delta^2 - \sigma^2}{\sigma^4} } \label{eqn:b2_o0}
\end{align}
where $H_n$ is the n-th probabilist's Hermite polynomial \citep[see also][]{Szalay_1988}. These estimators have already been derived and used in \citep{musso_2012, paranjape_2013a, paranjape_2013b} and we have extended them in \citetprob{} through scale dependent corrections, as we will review in Section \ref{sec:multivar_theory}.

In this paper, we will always measure the bias parameters independently of the model and we will then use the same values of ($b_1$, $b_2$) for comparing quadratic bias models with Gaussian bias models\footnote{Where for the Gaussian bias we additionally use that $\beta_1 = b_1$ and $\beta_2 = b_2 - b_1^2$.}. This is also the case for the bias models that we have shown in Figure \ref{fig:bias_function}, so that this is a fair comparison -- rather independently of the fitting method.

It is worth noting that the assumption of a Gaussian bias model for $f$ also results in a Gaussian for the galaxy environment distribution:
\begin{align}
    p_{\mathrm{gaus}}(\delta|g) &= p(\delta) f_{\mathrm{gaus}}(\delta) \\ 
              &= \frac{1}{\sqrt{2 \pi} \sigma_{\mathrm{g}}} \exp \left( - \frac{(\delta - \mu_{\mathrm{g}})^2}{2 \sigma_{\mathrm{g}}^2} \right)
\end{align}
where
\begin{align}
    \mu_{\mathrm{g}} &= \beta_1 \sigma^2 = \expectgal{\delta}\\
    \sigma_{\mathrm{g}}^2 &= \beta_2 \sigma^4 + \sigma^2 = \expectgal{(\delta - \mu_{\mathrm{g}})^2}
\end{align}
These parameters directly correspond to the mean and the variance of the galaxy environment distribution
.
Therefore, the selected Gaussian bias model also optimally fits the galaxy environment distribution. It is an intriguingly simple property of the Gaussian bias model, that all three, the bias function $f$, its large scale limit $F$ and the galaxy environment distribution $p(\delta|\mathrm{g})$ are Gaussian functions under this assumption.




\section{Measurements of the scale dependent density bias function}  \label{sec:measure_monovariate}

In this Section, we will show measurements of the scale dependent bias function $f$ for different setups and develop a variety of statistics to systematically test, how well different models approximate the function.

\subsection{Simulations}

Throughout this paper, we consider a single high-resolution cosmological simulation box. The simulation was run as part of the \quotes{BACCO simulation project} \citet{angulo_2021}. It has a boxsize of $L = 1440 h^{-1}\mathrm{Mpc}$ with $4320^3$ particles corresponding to a mass resolution of $m_p = \SI{3.2e9}{} h^{-1} M_\odot$. The cosmological parameters are $\Omega_m = 0.307$, $\Omega_\Lambda = 0.693$, $\Omega_b = 0.048$., $n_s = 0.9611$, $\sigma_8 = 0.9$, $h=0.677$ which are similar to the \citet{planck_2020} cosmology except for the roughly $10\%$ larger value of $\sigma_8$. To have a halo-mass versus bias relations similar to the more commonly used Planck cosmology, we use a snaphshot of the simulation at a scale-factor of $a=0.8$ where $\sigma_8$ has the effective value $D(a) \sigma_8 = 0.79$ where $D(a)$ is the linear growth factor normalized to $1$ at $a=1$.

\subsection{Tracer Catalogues}

To identify haloes the simulation code uses a modified version of \textsc{subfind}  \citep{Springel_2001} which first identifies haloes through a friends of friends (FoF) algorithm and subsequently calculates for each FoF group the mass $M_{200\mathrm{b}}$ in a region that encloses 200 times the mean density of the universe.

We consider haloes in narrow mass bins as tracers. For this we select all tracers that are in a $25\%$ range above and below a stated target mass
\begin{align}
    M_{200b} \in [M / 1.25, M \cdot 1.25 ] \,\,.
\end{align}
Halo selections throughout this article always use $a=0.8$, as explained above.

Further, we consider two catalogues of mock galaxies that are created based on subhalo abundance matching techniques in the same dark matter only simulation. The first set of ``stellar mass'' (SM) galaxies is optimized to mimic galaxies that may be observed in a survey like BOSS that use a cut in stellar mass. We choose a number density of $n = \SI{2e-3}{\h^3 \mega \parsec^{-3}}$, a redshift of $z \sim 1$ leading to a Lagrangian bias of order $b_1 \sim 0.5$. This catalogue is created with the SHAMe model that was introduced by \citet{Contreras_2021}. For a detailed discussion, we refer the reader to that paper. However, in short, this method uses an abundance matching technique based on the value of $v_{\mathrm{peak}}$ -- the highest circular velocity in the history of an object -- and uses additional prescriptions to model dynamical friction induced mergers and tidal stripping. The free parameters are tuned to mimic the clustering of stellar mass selected galaxies in the TNG300 hydrodynamical simulation \citep{nelson_2018, springel_2018, marinacci_2018, pillepich_2018, naiman_2018}.

As a second catalogue of galaxies, we consider ``star formation rate'' (SFR) selected galaxies as they would be observed by future surveys like EUCLID. For these we assume the same target parameters $n = \SI{2e-3}{\h^3 \mega \parsec^{-3}}$, a redshift of $z \sim 1$ and a Lagrangian bias of order $b_1 \sim 0.5$. However, for the abundance matching, we use a novel method developed by \citet{ortega_2024} named \quotes{SHAMe-SF}. This is an extension to the SHAMe method, in which additional adjustments have been made to accurately describe the redshift space clustering of star forming galaxies in the TNG300 simulations. We adopt again the set of parameters that optimally describes SFR selected galaxies from the TNG300. 

\subsection{Lagrangian fields} \label{sec:lag_fields}
\begin{figure}
    \centering
    \includegraphics[width=\columnwidth]{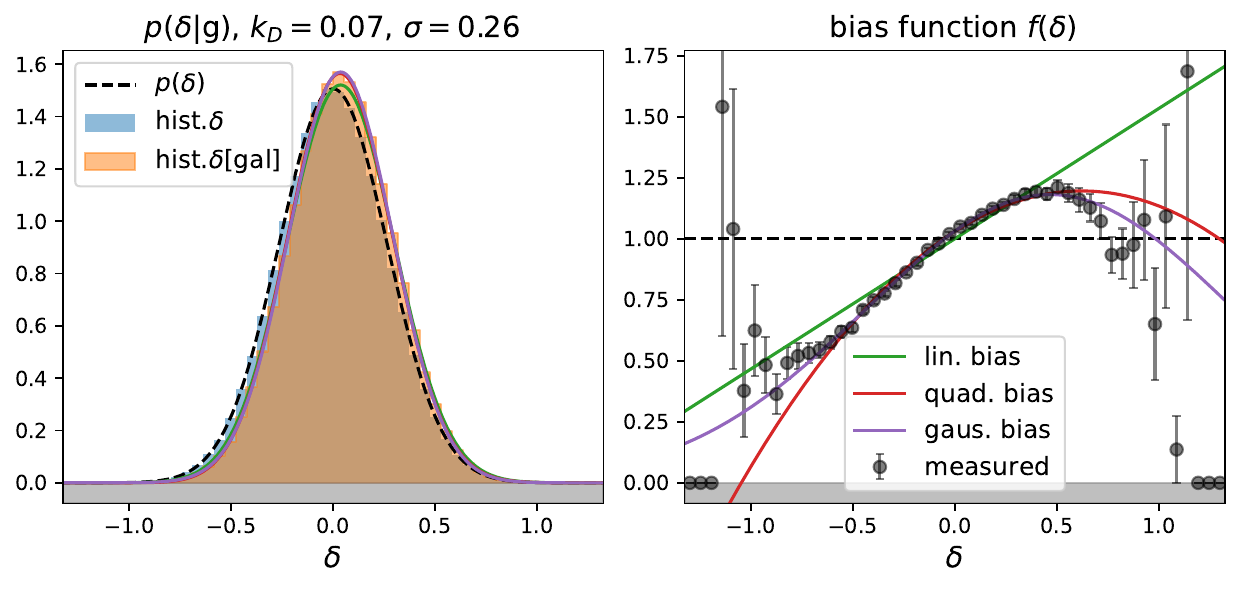} \\
    \includegraphics[width=\columnwidth]{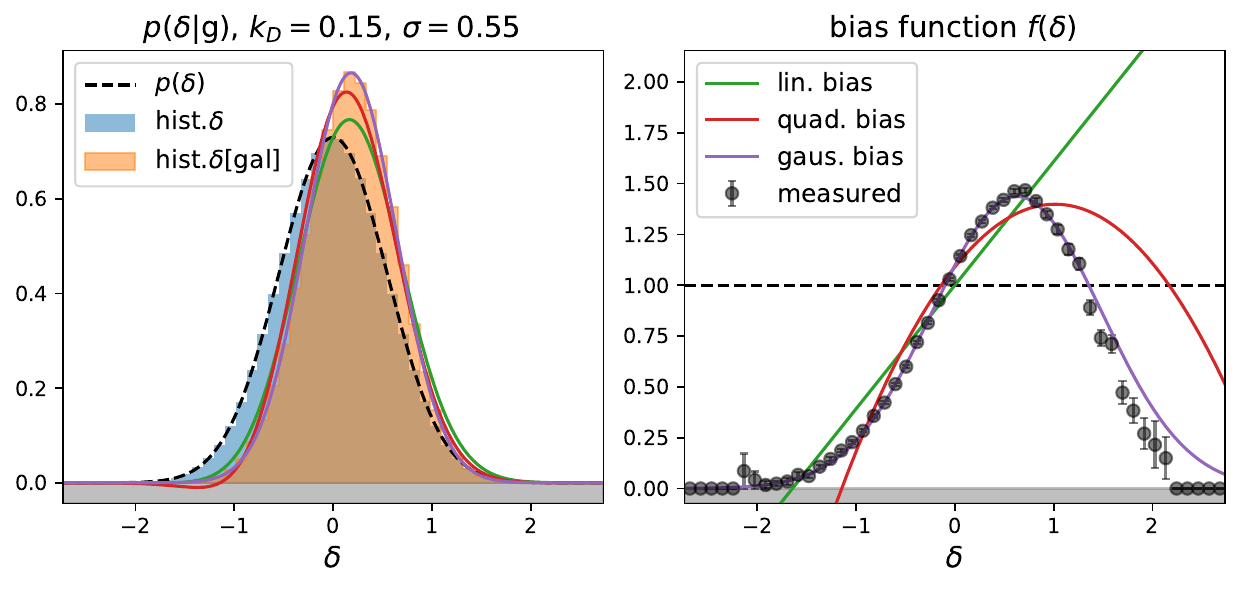} \\
    \includegraphics[width=\columnwidth]{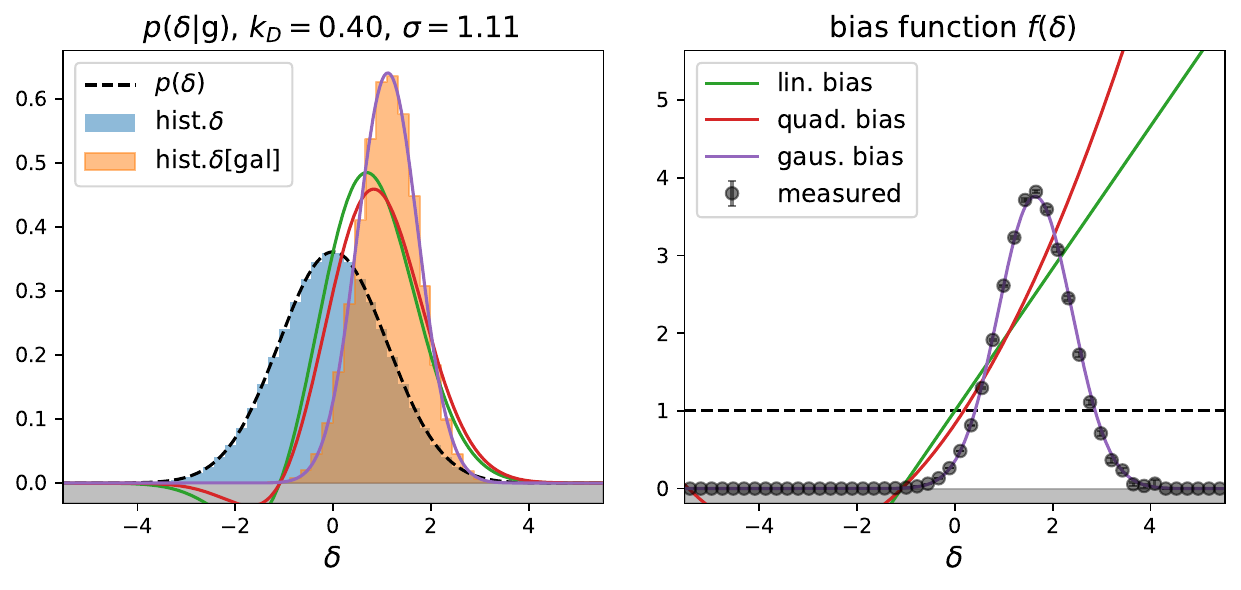} \\
    \caption{The galaxy environment distribution (left) and the Lagrangian bias function (right) of mass selected haloes ($M_{\rm{200b}} \sim \SI{4e13}{\per \h \msol}$) at different damping scales ($k_{\rm{d}} = \SI{0.07}{\h \per \mega \parsec}, \SI{0.15}{\h \per \mega \parsec}$ and $\SI{0.4}{\h \per \mega \parsec}$ in top, center and bottom respectively). At all scales the Gaussian bias provides a description that is as good as quadratic bias or significantly better -- especially so at small smoothing scales.}
    \label{fig:bias_functions_examples}
\end{figure}

\begin{figure}
    \centering
    \includegraphics[width=\columnwidth]{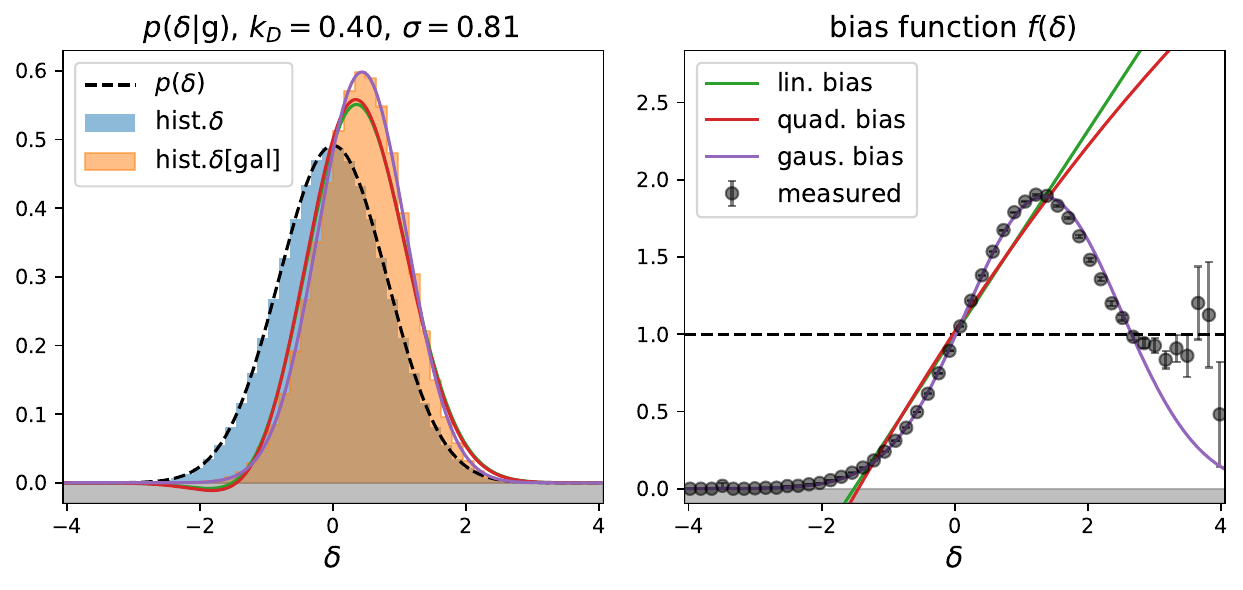}\\ \includegraphics[width=\columnwidth]{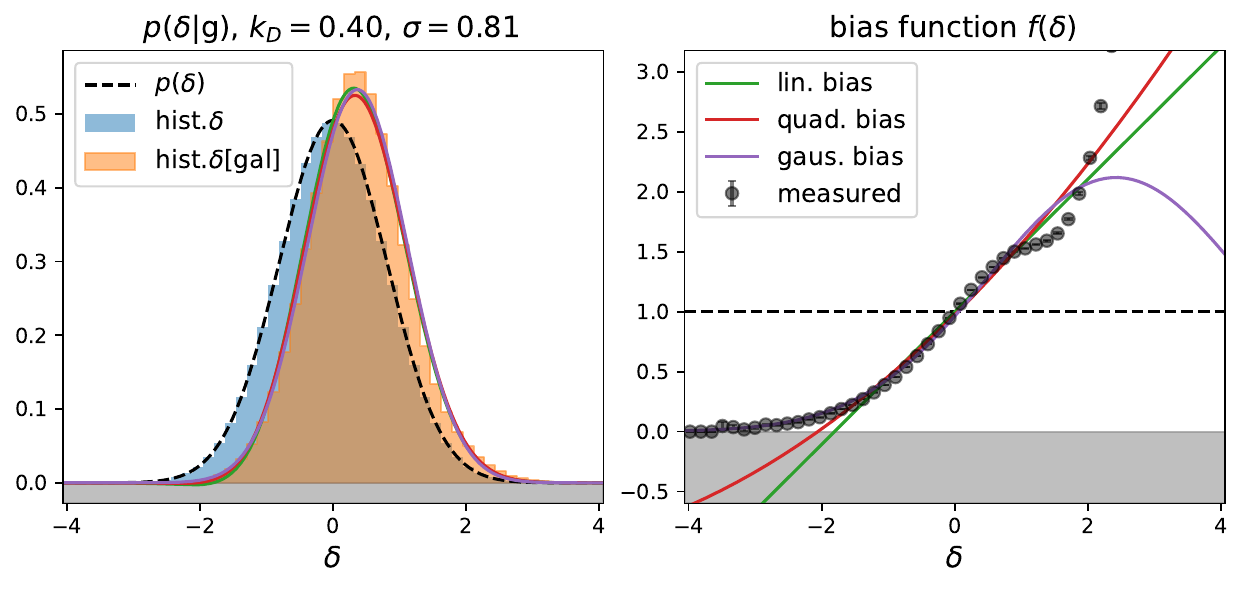}
    \caption{The galaxy environment distribution (left) and the Lagrangian bias function (right) of stellar mass selected galaxies (top) and star formation rate selected galaxies (bottom).}
    \label{fig:bias_functions_galaxies}
\end{figure}

To measure the galaxy environment distribution $p(\delta|\mathrm{g})$ and the bias function $f$, we need to know the linear field evaluated at the Lagrangian locations of the tracers. We approximate the Lagrangian location of each halo through the Lagrangian origin of its most bound particle. Since the simulation started from a Lagrangian grid, the Lagrangian origin of the most bound particle can easily be inferred from its id $i_{\mathrm{mb}}$ as 
\begin{align}
    \myvec{q}_{\mathrm{mb}} &= \frac{L}{N_{\mathrm{grid}}} \begin{pmatrix}
        i_x \\ i_y \\ i_z
    \end{pmatrix} \\
    i_{\mathrm{mb}} &= i_x N_{\mathrm{grid}}^2 + i_y N_{\mathrm{grid}} + i_z
\end{align}
where $N_{\mathrm{grid}} = 4320$ is the number of particles per dimension.

We know the linear density field of the simulation through the initial condition generator. To save computation time, we create a low resolution grid representation of the linear density field with $N_{\mathrm{lin}}^3$ grid points. For fields different than the linear density field we additionally multiply by the correct operator (e.g. $-k^2$ for the Laplacian) in Fourier space. We create a smoothed version of this field by multiplying with a sharp-k filter in Fourier space
\begin{align}
    \delta_k &= \delta_{\rm{lin},k} \cdot \Theta(k_{\rm{d}} - k)
\end{align}
with damping scale $k_{\rm{d}}$ where $\Theta$ is the Heavyside function.
We then deconvolve this field with a linear interpolation kernel and interpolate it to the Lagrangian locations of our tracer set. We choose $N_{\mathrm{lin}}$ sufficiently large that the resulting interpolated values are virtually independent of this discretization, e.g. $N_{\mathrm{lin}} = 183$ at $k_{\rm{d}} = 0.1 h^{-1}\mathrm{Mpc}$ and $N_{\mathrm{lin}} = 549$ at $k_{\rm{d}} = 0.3 h^{-1}\mathrm{Mpc}$.

Given the linear densities at the tracer locations, we can infer the galaxy environment distribution through a histogram of the densities and the bias function $f$ through a weighted histogram, where each tracer is weighted by $1/p(\delta)$. We always use 50 bins equally spaced in the range $\delta \in [-5\sigma, 5\sigma]$. We write the measurement of $f(\delta)$ as a vector $\myvec{f}$ where each component corresponds to the inferred value in a different bin. We then estimate the covariance matrix of the measurement through a Jackknife technique. For this we divide the box in Lagrangian space into $N_{\mathrm{jk}}^3$ sub-boxes with $N_{\mathrm{jk}} = 4$.  We perform $64$ measurements of $\myvec{f}_i$ by subsequently leaving out all tracers in one of the sub-boxes. Then we estimate the covariance matrix of the measurements 
\begin{align}
    \mat{C}_{\myvec{f}} &= \frac{1}{N_{\mathrm{jk}}^3 -1} \sum_i (\myvec{f}_i - \myvec{f}_0) \otimes (\myvec{f}_i - \myvec{f}_0) \\
    \myvec{f}_0 &= \frac{1}{N_{\mathrm{jk}}^3} \sum_i \myvec{f}_i.
\end{align}

Further, we estimate the bias parameters with the estimators from equations \eqref{eqn:b1_o0} and \eqref{eqn:b2_o0} and use these to parametrize the different bias models. The statistical uncertainties of these estimators are so small that we neglect them here.

\subsection{Measured Functions}

We show examples of bias function measurements for different scenarios in Figure \ref{fig:bias_functions_examples}. The first 3 panels use haloes of $M_{\rm{200b}} \sim \SI{4e13}{\per \h \msol}$  at different damping scales. At the largest scale $k_{\mathrm{d}} = \SI{0.07}{\h \per \mega \parsec}$, all bias models are indistinguishable and approximate the data well. At a smaller smoothing scale $k_{\rm{d}} = \SI{0.15}{\h \per \mega \parsec}$ small differences emerge and the Gaussian bias is a slightly better approximation than the quadratic bias. At very small damping scales $k_{\rm{d}} = \SI{0.4}{\h \per \mega \parsec}$ the expansion bias models completely fail to capture the data, whereas the Gaussian bias is still a very good approximation. Note that in all cases the bias models use the same renormalized bias parameters and the failure of expansion biases  cannot be attributed to fitting techniques. It seems generally that expansion bias struggles to capture scenarios where the bias function $f(\delta)$ gets already close to zero at intermediate density values $|\delta| / \sigma \lesssim 1$.

Beyond the accuracy of the models, we note a few relevant details in the scale dependence of the bias function. As explained in Section \ref{sec:gaussianbias}, under the assumption of a Gaussian density bias with scale-independent bias parameters, the location of the maximum of the bias function should be scale independent, but the width may be scale dependent. While the maximum of the bias function $f$ seems approximately at a consistent location between the two larger scales at $\delta_{\mathrm{max}} \approx 0.5$, at the smaller scale $k_{\rm{d}} = \SI{0.4}{\h \per \mega \parsec}$ the maximum has clearly shifted away to $\delta_{\mathrm{max}} \approx 1.5$. Such scale dependencies can be explained by the dependence on secondary variables, like for example the Laplacian, as we will discuss in Section \ref{sec:measure_multivariate} and Section \ref{sec:measureF}.

Finally, we consider the bias functions of the two mock galaxy catalogues. In the top panel of Figure \ref{fig:bias_functions_galaxies} we show the bias function of the stellar mass selected galaxy sample and in the bottom panel of the star formation rate selected galaxies, both at a damping scale $k_{\rm{d}} = \SI{0.4}{\h \per \mega \parsec}$. For the stellar mass selected galaxies the Gaussian is again much closer to the measured bias function than the expansion biases. For the star formation rate selected galaxies all models do not optimally approximate the bias function, especially at high densities. However, we might still favour the Gaussian approximation over the expansion in this scenario, since it approximates quite well, how the bias function approaches zero at small $\delta$. Despite these inaccuracies, it is worth noting that it is more relevant how well the bias models approximate the galaxy environment distribution $p(\delta|\mathrm{g})$, which is well recovered by all models.

We conclude that generally the Gaussian bias offers a much improved approximation to the bias function over a quadratic bias for many plausible scenarios. For scenarios where the Gaussian is not an optimal description, it still performs at a similar level to a quadratic bias function. We will show this systematically for a variety of metrics and tracer selections in the next section.

\subsection{Systematic evaluation} \label{sec:metrics1d}

\begin{figure}
    \centering
    \includegraphics[width=\columnwidth]{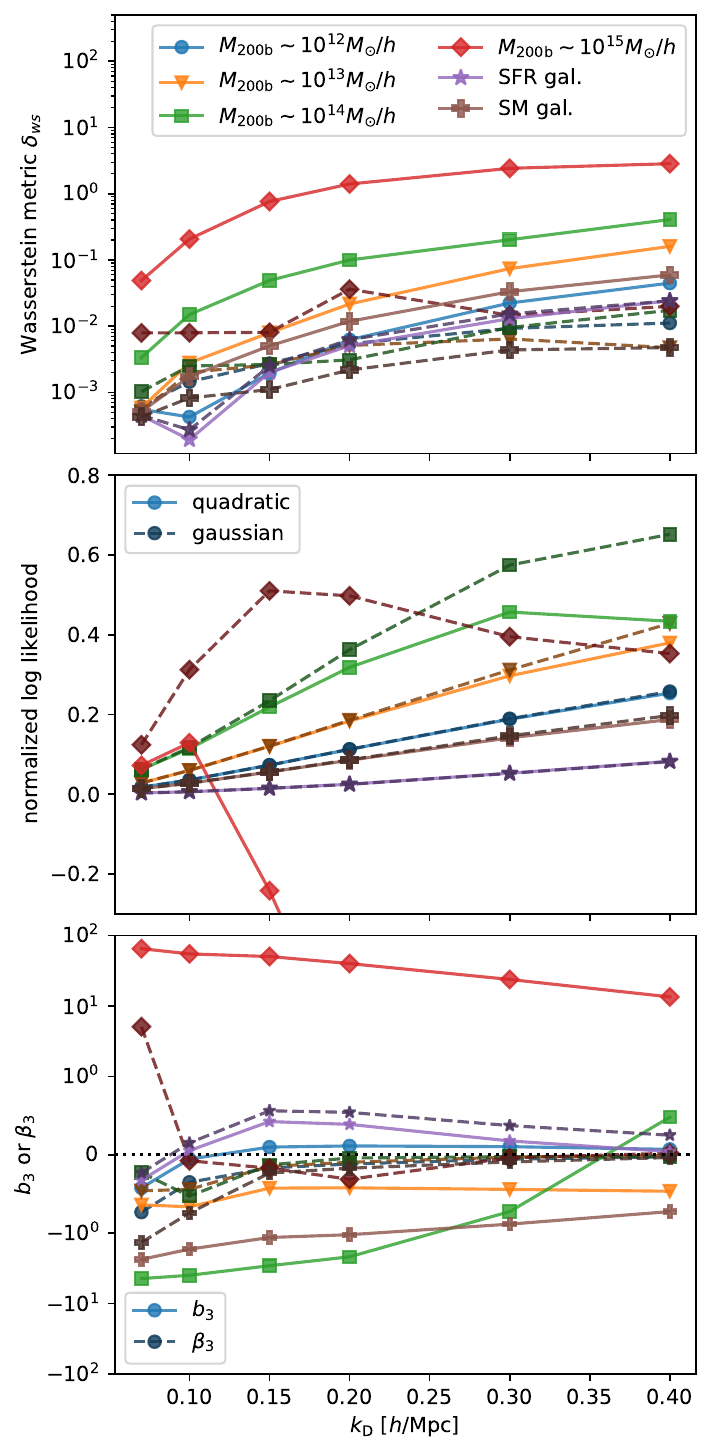}
    \caption{Metrics to compare the performance of the Gaussian bias model to expansion biases. The top panel shows the Wasserstein distance between the actual and the modelled galaxy environment distribution (smaller is better), the central panel the normalized log likelihood (larger is better) and the bottom panel the amplitude of the first unmodelled term (closer to zero is better). Brighter solid lines show the quadratic bias model and darker solid lines the Gaussian bias. All metrics and selections show the Gaussian bias performing either similar to a quadratic model or significantly better.}
    \label{fig:metrics1d}
\end{figure}

Here, we present metrics to systematically evaluate the performance of the considered bias models. The metrics are meant to compare the true galaxy environment distribution with the modelled one. It makes more sense, to compare the galaxy environment distributions than the bias functions, since this automatically weights the bias function appropriately by the Lagrangian volume. In principle, it would be desirable to use common probability theory measures for comparing the distributions, like for example the Kullback–Leibler divergence etc.. However, the expansion bias models generically predict distributions that are negatively valued for some values of $\delta$, so that the majority of probability metrics are not well defined. To alleviate this, we present as a first metric the Wasserstein metric, which is well defined even for negative values and as a second metric the likelihood with an appropriate method for clipping negative values.

The Wasserstein distance -- also known as the \quotes{Earth-movers} metric -- is the average absolute amount that the $\delta$-values of one distribution would need to be shifted to transform it into the other distribution. Conveniently this concept can also be applied for distributions that get negative so that it provides a fair comparison between the models. Further, it is simple to measure in one dimension and does not require coarse-graining the data (e.g. through a histogram). In one dimension, the Wasserstein distance between two distributions is given by
\begin{align}
    \delta_{ws} &= \int_0^1 \left|F^{-1}_1(F) - F^{-1}_2(F)\right| \mathrm{d} F
\end{align}
where $F^{-1}_1$ and $F^{-1}_2$ are the quantile functions -- i.e. the inverse of the cumulative distribution functions $F_1$ and $F_2$. However, since the measured area between the two curves is equivalent in $F$-space and in $\delta$-space, it can more conveniently be measured as
\begin{align}
    \delta_{ws} &= \int_{-\infty}^{\infty} \left|F_1(\delta) - F_2(\delta)\right| \mathrm{d} \delta  \,\,. \label{eqn:wasserstein_cdf}
\end{align}
For the negative distributions equation \eqref{eqn:wasserstein_cdf} can still be applied, but with a cumulative distribution function that is non-monotonic. For the discrete galaxy distribution, we approximate the cumulative distribution function with linear splines between the ordered data points ($\delta_n$, $n / N_{\rm{gal}}$) where $n$ is the rank in the ordered data points.

We show measurements of $\delta_{ws}$ in the top panel of Figure \ref{fig:metrics1d} evaluated  for halo selections of different masses and for the stellar mass and star formation rate selected galaxies at several different damping scales. The solid brighter lines correspond to quadratic bias model, whereas the darker dashed lines correspond to the Gaussian bias. Strikingly, $\delta_{ws}$ is quite small for the Gaussian bias model at all smoothing scales and for all tracer populations -- never getting beyond $\delta_{ws} \sim 0.03$. That means that using a Gaussian bias model a galaxy will be born at a smoothed linear density that is typically only off by $\delta_{ws} = 0.03$ or less. On the other hand, the performance of the quadratic bias model depends strongly on the selected set of tracers. It only behaves similarly well to the Gaussian bias for the SFR galaxies, but significantly worse than the Gaussian bias at large $k_{\mathrm{d}}$ for all other considered selections, easily differing on $\delta_{ws}$ by an order of magnitude, e.g. for SM galaxies or even two orders of magnitude for extremely massive haloes $M_{\rm{200b}} \sim \SI{e15}{\per \h \msol}$.

As the second metric, we consider the likelihood of observing a given set of environment densities ${\delta_i}$, given the model. This metric is only well defined for the Gaussian bias model, since only for that case is $p(\delta | g)$ a well defined (non-negative) probability density. Assuming that the linear densities $\delta_i$ of each tracer are randomly drawn from the distribution $p(\delta | g)$, the likelihood of observing the set of tracers is given by
\begin{align}
    L &= \prod_i^{N_{\rm{gal}}} p(\delta_i | g) \,\,.
\end{align}
To reduce the scaling with tracer numbers, we consider the log-likelihood per tracer. Further, the likelihood scales quite significantly with the width of the considered distribution. The log likelihood per tracer of a Gaussian that follows the background distribution would be given by
\begin{align}
    \expect{\log p(\delta)} &= \log(1/\sqrt{2 \pi \sigma^2}) - \expect{\frac{\delta^2}{2\sigma^2}} \nonumber \\
    &= -\frac{1}{2} \log(2 \pi \sigma^2) - \frac{1}{2} \,\,.
\end{align}
We subtract this term from the log likelihood to reduce the scaling with $\sigma$ and obtain a normalized log likelihood
\begin{align}
    \log L_n &= \expectgal{\log p(\delta | \mathrm{g}) } + \frac{1}{2} \log(2 \pi \sigma^2) + \frac{1}{2}
\end{align}
which should reach zero in the large scale limit $\sigma \rightarrow 0$ where the galaxy environment distribution approaches the background distribution.

To be able to evaluate this metric for the quadratic bias model we define a slightly modified distribution that is strictly positive, by clipping the bias function at a minimum value $f_{\mathrm{min}}$. The new distribution reads
\begin{align}
    p^*(\delta | g) &= N p(\delta) \begin{cases}
        f(\delta)  & \text{if } f(\delta) \geq f_{\mathrm{min}}  \\
        f_{\mathrm{min}} & \text{else}
    \end{cases} 
\end{align}
where the normalization constant $N$ is adjusted so that the area of $p^*$ is appropriately normalized to 1. By testing different clipping values $f_{\mathrm{min}} = \{ 0.001, 0.003, 0.01, 0.03, 0.1, 0.3 \}$ we have verified that the value of $ \log L_n$ is very insensitive to the choice of $f_{\mathrm{min}}$, except for the largest case $f_{\mathrm{min}}=0.3$ where it gets slightly reduced. However, for the comparison we will simply choose the best case scenario for the quadratic expansion, where we choose always the clipping value from the above choices that produces the largest likelihood at a given scale. For the Gaussian bias, we do not employ any clipping so that the clipping procedure can only enhance the likelihood of the expansion bias.


We show the normalized log-likelihood in the central panel of Figure \ref{fig:metrics1d}. The likelihood seems to be almost identical between the Gaussian and quadratic bias for low mass haloes $M_{\rm{200b}} \lesssim \SI{e12}{\per \h \msol}$ and for the SFR galaxies. For SM galaxies and intermediate mass haloes $\SI{e13}{\per \h \msol} \lesssim M_{\rm{200b}} \lesssim \SI{e14}{\per \h \msol}$ the Gaussian bias has a slightly improved log-likelihood per tracer. The differences are smaller at smaller $k_{\mathrm{d}}$ and larger at larger values of $k_{\mathrm{d}}$, which is expected, since for smaller values of $\sigma$ the bias function is mostly probed close to zero where the Taylor expansion is accurate. However, even small differences in the log likelihood per tracer may imply quite large differences in the full likelihood $L$ which scales with the number of tracers. Finally, we note that for the largest considered mass-selection $M_{\rm{200b}} \sim \SI{e15}{\per \h \msol}$ the Gaussian bias drastically outperforms the quadratic bias. As mentioned earlier, the likelihood is quite ill-defined for the quadratic bias in this case, but it is remarkable that such extreme cases are so well described by a Gaussian.


As a final measure of the difference of the true bias function with respect to the quadratic and Gaussian assumption, we consider the values of $b_3$ and $\beta_3$ respectively. The value of $b_3$ quantifies the lowest (third) order term that is not modelled in the quadratic model whereas the value of $\beta_3$ quantifies the lowest order term that is not modelled by the Gaussian. Recall that the Gaussian implicitly models some aspects of $b_3$ through its implicit dependence on $b_1$ and $b_2$, as can be seen from equation \eqref{eqn:beta3_of_b} under the assumption $\beta_3=0$.

We show the obtained values of $b_3$ and $\beta_3$ in the bottom panel of Figure \ref{fig:metrics1d}. First of all, we note that these parameters are quite scale dependent which is due to neglecting the Laplacian bias, as we discuss in detail in \citetprob. However, these parameters still describe, how well the third moment is captured by the bias models at the considered scale. Clearly $\beta_3$ is very small, with $|\beta_3| < 1$ for almost all considered cases and generally getting even smaller for large $k_{\mathrm{d}}$.  On the other hand we see that for some selections $b_3$ can be very large, for example $b_3 \gg 10$ for the largest mass haloes, indicating a catastrophic behaviour of the canonical bias expansion for such large mass objects or $\beta_3 \sim -5$ for $M_{\rm{200b}} \sim \SI{e14}{\per \h \msol}$ at $k_{\mathrm{d}} \lesssim \SI{0.2}{\h \per \mega \parsec}$. See \citetprob, for a more detailed discussion of the behaviour of $b_3$ versus $\beta_3$.

We conclude that in all cases the Gaussian bias model approximates the density bias function at least as well as a quadratic bias model, and for most cases it does so significantly better. Note-worthily, the Gaussian bias still poses a good approximation for scenarios where the canonical expansion is typically expected to break down, for example for very high mass haloes or for very small smoothing scales. 

While this shows that a Gaussian is always a good description for the galaxy density-environment distribution at a given scale, this does not yet show that it is always sufficient to fully describe all aspects of the galaxy field. For this it is also necessary to consider the ability of the model to jointly describe all scales that are larger than the considered smoothing scale and to show that other variables, like e.g. the Laplacian and the tidal field, can appropriately be incorporated. We will make the necessary theoretical considerations in the next section and we will evaluate it quantitatively in Sections \ref{sec:measure_multivariate} and \ref{sec:measureF}.

\section{The multivariate bias function} \label{sec:multivar_theory}
Galaxy formation does not only get affected by a uniform large scale density, but it can also get affected by other aspects of the environment, such as the tidal field or the Laplacian of the density field
\begin{align}
    L &= \nabla^2 \delta \,\,.
\end{align}
Therefore, multiple variables may be necessary to fully characterize the bias function. See e.g. \citet{bartlett_2024} for a clear demonstration that local density-only bias models (in Eulerian space) are already an insufficient description of the halo density field on fairly large scales. 

In this Section, we generalize the concepts introduced in Section \ref{sec:theory} to the multivariate case. Additionally, we show how to reconstruct the large scale bias function $F$ from measurements at finite damping scales in Section \ref{sec:measuringF}. Further, in Section \ref{sec:gaussiantidal}, we will briefly mention how a dependence on the tidal field can be included into a Gaussian bias model (though we will not use it in this article).

\subsection{Definitions}
We will characterize the linear field through a vector, e.g. $\myvec{x} = (\delta, L)^T$, that contains all scalar variables of interest. Our examples will mainly focus on the joint treatment of density and Laplacian, but the considerations in this Section generalize to other (linear) variables as well.

Most equations that were derived in Section \ref{sec:theory} generalize in a straightforward way to the multivariate case. The large scale bias function $F$ and the scale dependent bias function $f$ are again related through
\begin{align}
    F(\myvec{x}_0) &= \expect{f(\myvec{x} + \myvec{x}_0)} \label{eqn:mv_renorm}
\end{align}
where the average now goes over a multivariate Gaussian distribution
\begin{align}
    p(\myvec{x}) &= \frac{1}{2 \pi \sqrt{\det(\mat{C})}} \exp \left(- \frac{1}{2} \myvec{x}^T \mat{C}^{-1} \myvec{x}  \right) \label{eqn:multivariate_gaussian}
\end{align}
where $\mat{C}$ is the covariance matrix between the components of $\myvec{x}$. For example, for the joint distribution of density and Laplacian, we have
\begin{align}
\mat{C}_{\delta,L} &= \left[\begin{matrix}\sigma_{0}^{2} & - \sigma_{1}^{2}\\- \sigma_{1}^{2} & \sigma_{2}^{2}\end{matrix}\right]
\end{align}
where $\sigma_0^2 = \expect{\delta^2} = \sigma^2$, $\sigma_2^2 = \expect{L^2}$ and $\sigma_1^2 = \expect{-\delta L}$.

The bias parameter $\myvec{b}_1$ now becomes a vector and $\mat{b}_2$ a matrix:
\begin{align}
    \myvec{b}_1 &= \left. \nabla_{\myvec{x}} F \, \right|_{\myvec{x} = 0} \\ %
    \mat{b}_2 &= \left. (\nabla_{\myvec{x}} \otimes \nabla_{\myvec{x}}) F \, \right|_{\myvec{x} = 0}
\end{align}
where \quotes{$\otimes$} denotes an outer product. For the case $\myvec{x} = (\delta, L)^T$ the components of $\myvec{b}_1$ correspond to the linear density bias parameter, which we will label \quotes{$b_\delta$} in this section to avoid confusion (before called \quotes{$b_1$}), and the Laplacian bias $b_L$. The components of $\mat{b}_2$ correspond to second derivative biases, including $b_{\delta \delta}$  (before called \quotes{$b_2$}), $b_{\delta L}$ and $b_{LL}$. 

Note that the terms $b_{\delta L}$ and $b_{LL}$ are often considered to be of higher order than 2 in perturbation theory. How orders are assigned to different terms depends on the assumption of how spatial derivatives are discounted versus the appearance of higher order powers of the same variable in the expansion. For mathematical simplicity, we will only count orders by the powers of the variables here so that e.g. the term $\delta L$ is counted as second order.

\subsection{Multivariate Gaussian Bias}
We again define cumulant biases through the derivatives of $\log F$:
\begin{align}
    \myvec{\beta}_1 &= \left. \nabla_{\myvec{x}} \log F \, \right|_{\myvec{x} = 0} = b_1 \\ %
    \mat{\beta}_2 &=  \left. (\nabla_{\myvec{x}} \otimes \nabla_{\myvec{x}}) F \, \right|_{\myvec{x} = 0} = \mat{b}_2 - \myvec{b}_1 \otimes \myvec{b}_1
\end{align}
and write the large scale bias function
\begin{align}
    \log F &= \myvec{\beta}_1 \myvec{x} + \frac{1}{2} \myvec{x}^T \mat{\beta}_2 \myvec{x}
\end{align}
and assume a multivariate Gaussian bias model
\begin{align}
    f(\myvec{x}) &= N_b \exp \left(- \frac{1}{2} (\myvec{x} - \myvec{\mu}_b)^T \mat{C}_{b}^{-1} (\myvec{x} - \myvec{\mu}_b)  \right) \,\,. \label{eqn:multiv_gaussian_biasfunc}
\end{align}
By applying equation \eqref{eqn:mv_renorm}, we find the coefficients
\begin{align}
    \myvec{\mu}_b  &=  -\beta_2^{-1} \myvec{\beta}_1 \\
    \mat{C}_b      &=  -\beta_2^{-1} - \mat{C} \\
        N_b        &= \frac{\exp \left(- \frac{1}{2} \myvec{\beta}_1^T \beta_2^{-1} \myvec{\beta}_1 \right)}{\sqrt{\det (\mat{1} + \mat{C} \beta_2 )}}
\end{align}
which are simple generalizations of the results from Section \ref{sec:theory}. 

We note that the multivariate Gaussian bias of two variables has by default five free parameters -- two describing the location of the maximum and three describing the covariance matrix. 

\subsection{Multivariate Expansion}
The renormalized form of the quadratic expansion model can then be written as
\begin{align}
    f(\myvec{x}_0) &= 1 + \myvec{b}_1 \myvec{x} + \frac{1}{2} (\myvec{x}^T \mat{b}_2 \myvec{x} - \trace{\mat{C} \mat{b}_2}) \\
                   &= 1 + b_{\delta} \delta + b_L L + \frac{1}{2} b_{\delta \delta} (\delta^2 - \sigma_0^2) \nonumber \\
                   &\text{\quad} + b_{\delta L} (\delta L + \sigma_1^2) + \frac{1}{2} b_{LL} (L^2 - \sigma_2^2)) \,\,.
\end{align}
The last two terms are not considered as second order terms in most studies, because terms that include spatial derivatives (like the Laplacian $L$) are often counted as higher order terms than those which just include $\delta$. However, since it is most natural to include the corresponding terms in the multivariate Gaussian bias model, we will here use this five parameter model as the \quotes{multivariate bias expansion model} as a fair comparison with the Gaussian bias.

\subsection{Measuring bias parameters}
Just as in Section \ref{sec:theory}, we can measure the multivariate bias function $f$ by measuring the galaxy environment distribution $p(\myvec{x} | \mathrm{g})$ through a multivariate histogram and dividing by $p(\myvec{x})$.

Further, we can measure bias parameters through the same method as in Section \ref{sec:measure_monovariate} -- that is through the moments of the galaxy environment distribution. As derived in \citetprob, the bias parameters can be measured through
\begin{align}
    \myvec{b}_1 &= \expectgal{\mat{C}^{-1} \myvec{x} } \label{eqn:measure_b1vec} \\
    \mat{b}_2 &= \expectgal{\mat{C}^{-1} (\myvec{x} \otimes \myvec{x} - \mat{C}) \mat{C}^{-1} } \,\,. \label{eqn:measure_b2mat}
\end{align}
Note that if we consider $\myvec{x} = (\delta, L)$, the obtained density bias estimator is different to the pure density case from equation \eqref{eqn:b1_o0}:
\begin{align}
    b_{\delta} &= \expectgal{\frac{\delta \sigma_{2}^{2} + L \sigma_{1}^{2}}{\sigma_{0}^{2} \sigma_{2}^{2} - \sigma_{1}^{4}}} \,\,. \label{eqn:b1_o2}
\end{align}
As we investigate in large detail in \citetprob, this estimator is much more independent of the scale it is evaluated at than equation \eqref{eqn:b1_o0}, since the Laplacian term corrects for the scale dependence. 

We note again that for the Gaussian bias model the galaxy environment distribution $p(\myvec{x}| \mathrm{g}) = p(\myvec{x}) f(\myvec{x})$ is a (conventionally normalized) multivariate Gaussian and that our fitting method ensures that its mean and covariance matrix correspond exactly to the mean and covariance of the galaxy environment distribution
\begin{align}
    \myvec{\mu}_{\mathrm{g}} &=  \expectgal{\myvec{x}} \\
    \mat{C}_{\mathrm{g}} &=  \expectgal{(\myvec{x} - \myvec{\mu}_{\mathrm{g}}) \otimes (\myvec{x} - \myvec{\mu}_{\mathrm{g}})} \,\,.
\end{align}
Arguably, the simplest way to fit and parameterize the multivariate Gaussian bias model is by inferring the mean and covariance of the galaxy environment distribution as above and then write $f$ through the ratio $p(\myvec{x}| \mathrm{g}) / p(\myvec{x})$. However, the explicit representation through bias parameters is still useful to predict the scale dependence and to compare it fairly with an expansion bias model.

\subsection{Measuring the large scale bias function} \label{sec:measuringF}
It is not only possible to measure $f$ in a non-parametric way, but we can also directly reconstruct the large scale bias function $F$ at finite damping scales:
\begin{align}
    F(\myvec{x}_0) &=  \langle f(\myvec{x} + \myvec{x}_0) \rangle \nonumber \\
        &= \int_{-\infty}^\infty  p(\myvec{x}) f(\myvec{x} + \myvec{x}_0) \mathrm{d} \myvec{x}  \nonumber \\
        &= \int_{-\infty}^\infty  p(\myvec{x} - \myvec{x}_0) f(\myvec{x}) \mathrm{d} \myvec{x}  \label{eqn:ng_from_pf} \,\,.
\end{align}
Therefore, we could in principle obtain $F$ by first measuring $f$ and then convolving it with a Gaussian. However, a more accurate method to measure $F$ that requires fewer discretization steps is the following: 

For the multivariate background distribution it is
\begin{align}
    p(\myvec{x} - \myvec{x}_0) 
    &= \frac{1}{2 \pi \sqrt{\det(\mat{C})}} \left(- \frac{1}{2} \myvec{x}^T \mat{C}^{-1} \myvec{x} + \myvec{x}^T \mat{C}^{-1} \myvec{x}_0 - \frac{1}{2} \myvec{x}_0^T \mat{C}^{-1} \myvec{x}_0 \right) \nonumber \\
    &= p(\myvec{x}) \exp \left(\myvec{x}^T \mat{C}^{-1} \myvec{x}_0 \right) \exp \left(- \frac{1}{2} \myvec{x}_0 ^T \mat{C}^{-1} \myvec{x}_0 \right) \nonumber
\end{align}
where we have used the symmetry of the covariance matrix $\myvec{x}^T \mat{C}^{-1} \myvec{x}_0 = \myvec{x}_0^T \mat{C}^{-1} \myvec{x}$. Therefore, we find
\begin{align}
    F(\myvec{x}_0)  &= \exp \left(- \frac{1}{2} \myvec{x}_0 ^T \mat{C}^{-1} \myvec{x}_0 \right) \int_{-\infty}^\infty  p(\myvec{x}|\mathrm{g}) \exp \left(\myvec{x}^T \mat{C}^{-1} \myvec{x}_0 \right) \mathrm{d} \myvec{x} \nonumber \\
                    &= \exp \left(- \frac{1}{2} \myvec{x}_0 ^T \mat{C}^{-1} \myvec{x}_0 \right) \expectgal{\exp \left(\myvec{x}^T \mat{C}^{-1} \myvec{x}_0 \right)} \,\,. \label{eqn:measureF}
\end{align}
Note that the appearing expectation value corresponds to a moment generating function \citepprob{}. If we consider only the monovariate distribution of densities $\myvec{x} = \delta$, we have
\begin{align}
    F(\delta_0) &= \exp \left(- \frac{\delta_0^2}{2 \sigma^2}  \right) \expectgal{\exp \left(\frac{\delta \delta_0}{\sigma^2} \right)} \,\,. \label{eqn:measureFo0}
\end{align}
We call this the \quotes{spatial order 0} estimator of the renormalized density bias function.
However, if we consider the joint distribution of density and Laplacian,
 we get a different estimator, even for pure density displacements $\myvec{x}_0 = (\delta_0, 0)^T$
\begin{align}
    F(\delta_0) &= \exp \left(- \frac{\delta_0^2 \sigma_2^2}{2 \sigma_*^4}  \right) \expectgal{\exp \left(\delta_0 \frac{ \delta \sigma_2^2 + L \sigma_1^2 }{\sigma_*^4} \right)} \label{eqn:measureFo2}
\end{align}
which we call the \quotes{spatial order 2} estimator of the renormalized density bias function, since it contains corrections based on the Laplacian of the density field. 

We want to emphasize here the important observation that the renormalized bias function is not only an abstract theoretical concept, but it can also directly be measured in simulations. Further, it means that it is possible to define non-parametric bias approaches at some smoothing scale and to rescale them to different smoothing scales, e.g. by directly discretizing $F$.

While we will focus in Section \ref{sec:measure_multivariate} on measurements of the scale dependent bias function $f$, we will show reconstructions of the scale-independent large scale bias function in Section \ref{sec:measureF}. Whether both $f$ and $F$ can be consistently described through the same bias model can be used as a test of whether a bias model accurately captures the deterministic dependencies on scales larger than the damping scale. Further, whether the non-parametric measurement $F$ is consistent across different smoothing scales can be used as a test of the range of validity of the PBS assumption for a given set of variables -- independently of any assumptions about the functional form.

\subsection{Gaussian Tidal bias} \label{sec:gaussiantidal}
For the sake of simplicity, and since the tidal bias of haloes is generally very low, we will not consider tidal fields for the measurements in the remainder of this paper. However, since it is very common to include the tidal field in a bias expansion at second order, we want to briefly mention how it can be included in the Gaussian bias framework for the benefit of future studies. We define the traceless tidal tensor
\begin{align}
    \mat{K} &= (\myvec{\nabla} \otimes \myvec{\nabla}) \phi - \frac{\delta}{3} \mat{J}_2
\end{align}
where $\mat{J}_2$ is the unit matrix. We show in Appendix \ref{app:renormtidal} that (at order two) the joint bias function of $(\delta, L, \mat{K})$ factorizes as
\begin{align}
    f(\delta, L, \mat{K}) &= f_0(\delta, L) f_{\mat{K}}(\mat{K})
\end{align}
where the bias function of density and Laplacian $f_0$ is given as earlier by equation \eqref{eqn:multiv_gaussian_biasfunc} for $\myvec{x} = (\delta, L)^T$ and the tidal component of the bias function is given by
\begin{align}
    f_{\mat{K}}(\mat{K}) &= \left|1 + \frac{4 \sigma^2}{15} b_{K^2} \right|^{-5/2} \exp \left(\frac{b_{K^2} K^2}{1 + \frac{4}{15} \sigma^2 b_{K^2}} \right) \label{eqn:fk_renorm}
\end{align}
where $K^2 = \trace{\mat{K} \mat{K}}$ and $f_{\mat{K}}$ has the large scale limit
\begin{align}
    F_{\mat{K}}(\mat{K}_0) &= \expect{f_{\mat{K}}(\mat{K} + \mat{K}_0)} \nonumber \\
                 &= \exp \left(b_{K^2} K_0^2  \right) \,\,.
\end{align}
The form of $f_{\mat{K}}$ can be understood as follows: The normalization factor originates from the normalization of a five dimensional Gaussian, since the distribution $p(\mat{K})$ is effectively five-dimensional. Further, the distribution of the tidal tensor has per degree of freedom the variance
\begin{align}
    \frac{1}{5} \langle K^2 \rangle &= \frac{2\sigma^2}{15} 
\end{align}
which naturally appears in equation \eqref{eqn:fk_renorm}. 

We summarize that it is possible to phrase a renormalized Gaussian bias model in the multivariate case -- e.g. using the density, Laplacian and tidal field as variables. Therefore, it is possible to encode all terms that are traditionally considered in a second order expansion in a fully self consistent Gaussian bias model that is strictly positive and for which $f(\delta, L, \mat{K})$, $F(\delta, L, \mat{K})$ and $p(\delta, L, \mat{K}| \mathrm{g})$ are all simple multivariate Gaussians.

\section{Measurements of the Multivariate scale dependent bias function} \label{sec:measure_multivariate}
In this section, we will show measurements of the multivariate bias function considering the density and the Laplacian of the density field. Here, we will only check whether the models describe the bias relation at a single scale, whereas we will consider the problem across multiple scales in Section \ref{sec:measureF}.

\subsection{Example Function}
\begin{figure}
    \centering
    \includegraphics[width=\columnwidth]{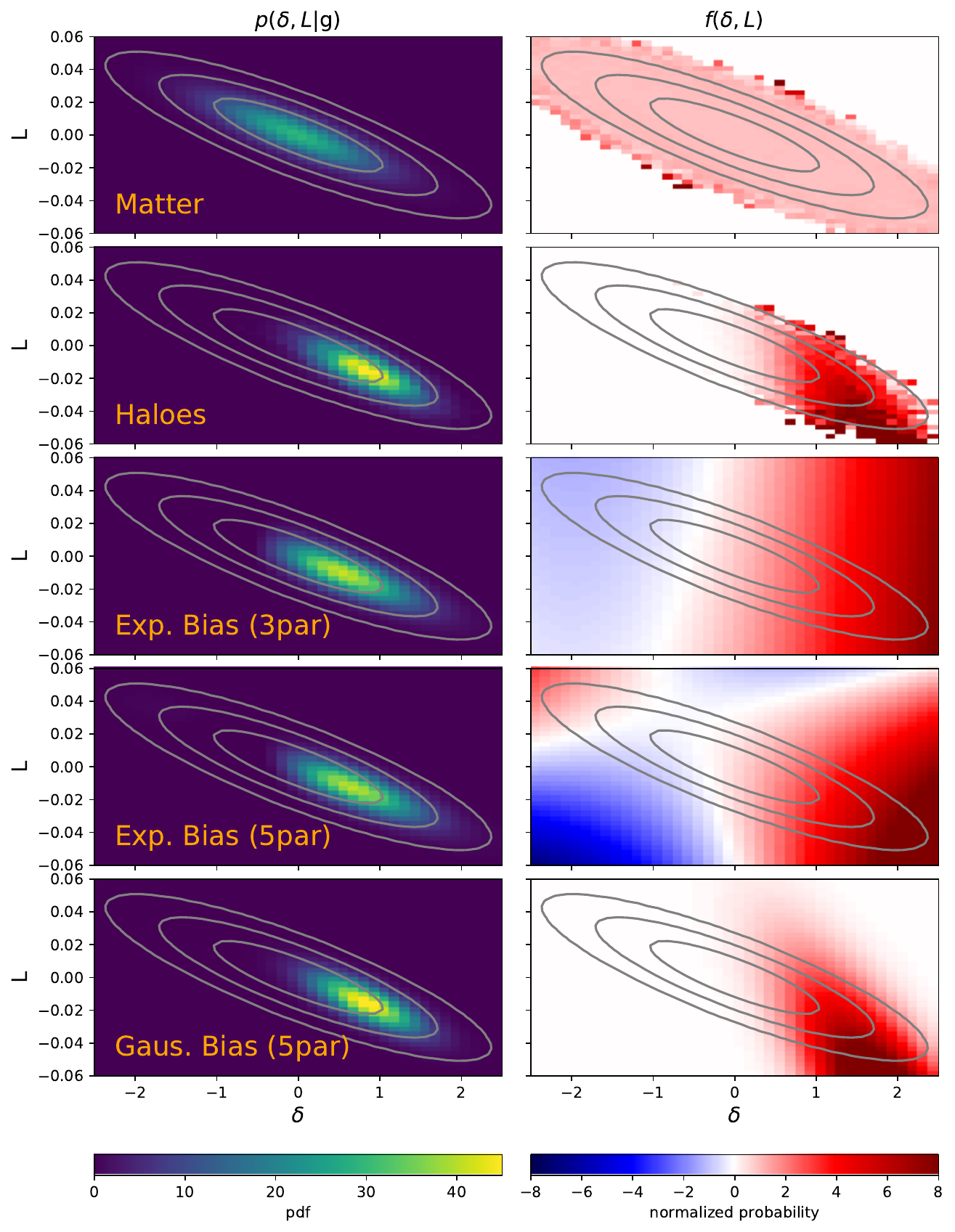}
    \caption{The distribution of Lagrangian over-densities $\delta$ and the Laplacian $L$ (in units of $h^2 \mathrm{Mpc}^{-2}$)  $p(\delta, L | g)$ (left) and the multivariate bias function (right).  The top row shows the distribution for matter particles for $k_{\rm{d}} = \SI{0.2}{\h \per \mega \parsec}$ and the second row of haloes selected in the mass-range $M_{\rm{200b}} \sim \SI{2e14}{\per \h \msol}$. The remaining rows show approximations to the halo distribution through a 3 parameter second order expansion, a 5 parameter second order expansion and a Gaussian bias model. The contours show the 1-,2- and 3-sigma regions of the matter distribution. The measurement for haloes has only reasonable statistics roughly inside of the 3-sigma region of the matter distribution (outermost grey contour). The multivariate Gaussian bias seems to be a good approximation to the actual distribution and adequately respects the positivity constraint}
    \label{fig:multivariate_biasfunc}
\end{figure}

First, we visualize the multivariate bias function for one example case. Here we select haloes with masses $M_{\rm{200b}} \sim \SI{2e14}{\per \h \msol}$ at a damping scale $k_{\rm{d}} = 0.2 h^{-1}\mathrm{Mpc}$. We measure the galaxy environment distribution $p(\delta, L | \mathrm{g})$ through a normalized histogram of ($\delta$, $L$) at the locations of galaxies and we measure the bias function through the histogram weighted by $1/p(\delta,L)$. 

In the left column of Figure \ref{fig:multivariate_biasfunc}, we show the galaxy environment distribution for different cases whereas the right panels show the bias function for the respective cases. The matter distribution is -- as expected -- a multivariate Gaussian and leads to a bias function with $f=1$ in the measured regime. 
Considering the distribution of haloes in the second row of Figure \ref{fig:multivariate_biasfunc}, we find that the bias function is measured reasonably well within the 3$\sigma$ region (outermost grey contour). The distribution of haloes is offset from the background distribution, clearly preferring higher densities and more negative values of the Laplacian.

\begin{figure}
    \centering
    \includegraphics[width=\columnwidth]{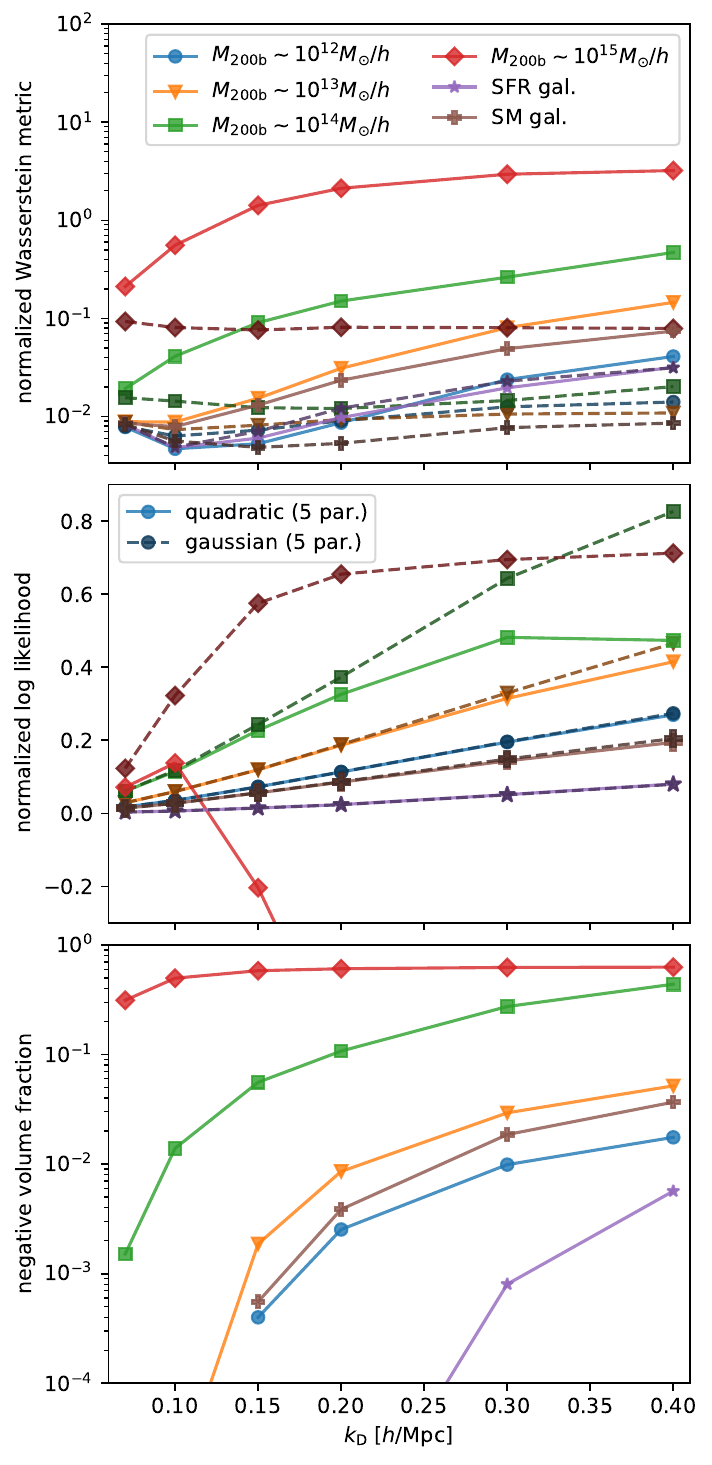}
    \caption{Metrics for evaluating the performance of the different bias models in the multivariate case. The top panel shows the Wasserstein metric (smaller is better), the central panel the normalized log likelihood (larger is better) and the bottom panel the fraction of volume for which negative galaxy densities are predicted. Generally, the multivariate Gaussian bias model appears to describe the data either significantly better or at least equally well to a quadratic bias model and it never predicts negative galaxy densities, which can get quite significant for the expansion bias. }
    \label{fig:mv_metrics}
\end{figure}

We consider three different approximations to the bias function -- a quadratic expansion bias with $b_{\delta}, b_{\delta,\delta}$ and $b_{L}$ as parameters, a quadratic bias model with the five parameters $b_{\delta}, b_{\delta,\delta}$, $b_{L}$, $b_{\delta,L}$, $b_{L,L}$, and a Gaussian bias with the same parameters. All of these parameters are inferred as in equations \eqref{eqn:measure_b1vec} and \eqref{eqn:measure_b2mat} (through moments of the galaxy environment distribution) independently of the considered model so that this is a fair comparison independent of fitting technique. While the quadratic biases both capture the rough shape of the environment distribution and of the bias function reasonably well, they also have some severe shortcomings. Most significantly they predict negative values for significant fraction of space. On the other hand, the Gaussian bias is as expected strictly positive and it describes the shape of the actual bias function almost perfectly. It is quite plausible that the Gaussian would also be a significantly better description of the bias function in this scenario than expansions of significantly higher order. The built-in positivity constraint makes it significantly easier to approximate the actual bias function which is very close to zero over a significant fraction of space.

\subsection{Systematic Evaluation} 
Here, we will systematically evaluate the performance of the quadratic versus Gaussian bias model for the same set of damping scales and tracer selections as in Section \ref{sec:metrics1d}.

We again consider the Wasserstein distance between the two distributions. In the two dimensional scenario it is necessary to assume a metric between $\delta$ and $L$. The most natural choice is the distance notion
\begin{align}
    d(\myvec{x}_1, \myvec{x}_2) &= \sqrt{(\myvec{x}_1 - \myvec{x}_2)^T C^{-1}_{\delta, L} (\myvec{x}_1 - \myvec{x}_2)}
\end{align}
which measures the distance in units of standard deviations. The Wasserstein distance is then again the minimal average amount that probability mass has to be transported to transform the first distribution into the second. In two dimensions calculating the Wasserstein distance is a rather complicated optimization problem. We use the \textsc{POT} (Python Optimal Transport) library to estimate the Wasserstein distance between the actual and the measured distribution. To obtain a manageable computation time, we discretize the distributions to a grid that covers the 5$\sigma$-region (in the principal component frame) through $50^2$ bins. We have checked that increasing the region or the number of bins does not affect the results notably. Since the Wasserstein distance is invariant under adding constant offsets to both distributions, it is also well defined for the case with negative probability densities.

We show the thus obtained Wasserstein distances in the top panel of Figure \ref{fig:mv_metrics}. The overall picture is very similar to what we have found in  Section \ref{sec:measure_monovariate}: The multivariate Gaussian bias has always smaller or equal values of the Wasserstein distance when compared to the quadratic (five parmeter) expansion. The difference is most significant for high mass, high bias objects $M_{200\rm{b}} \gtrsim \SI{e14}{\per \h \msol}$ reaching more than order of magnitude at large $k_{\mathrm{d}}$. For intermediate mass objects and stellar mass selected galaxies the difference is still quite significantly smaller for the Gaussian bias. For stellar mass selected galaxies the Wasserstein metric is almost equal between the Gaussian and the expansion bias, though being even slightly smaller for the expansion bias at some intermediate scales, e.g. at $k_{\mathrm{d}} = 0.3 h \mathrm{Mpc}^{-1}$.

As a secondary metric we again consider the log-likelihood, with the likelihood now normalized to the multivariate Gaussian case:
\begin{align}
    \log L_n &= \expectgal{\log p(\delta, L | \mathrm{g}) } + \frac{1}{2} \log(4 \pi^2 \det{C}) + 1 \,\,.
\end{align}
For the quadratic bias we use the same clipping procedure as described in Section \ref{sec:metrics1d}. We have again verified that results are quite independent of the clipping value, and we always use the one that leads to the largest likelihood.

We show the likelihood values in the second panel of Figure \ref{fig:mv_metrics}. The picture is very similar to the monovariate case from Figure \ref{fig:metrics1d} with some differences being additionally enhanced.

Finally, we measure the fraction of Lagrangian Volume that is assigned a negative probability (and therefore galaxy density) by the expansion bias
\begin{align}
    V_{\mathrm{neg}} &= 1 - \expect{H(f(\myvec{x}))}
\end{align}
where $H$ is the Heaviside step function. We show the negative volume fraction in the bottom panel of Figure \ref{fig:mv_metrics}. While the Gaussian bias has always $V_{\mathrm{neg}} = 0$, the expansion bias predicts varying amounts of negative volume. Generally, the problem gets worse with larger $k_{\mathrm{d}}$ and with stronger biased tracers. E.g. for the SFR galaxies the fraction seems to be below $1\%$ at all scales, but about half of the volume is predicted to be negative for the haloes with $M_{\mathrm{200b}} \sim 10^{14} M_{\odot}$ at the largest $k_{\mathrm{d}}$.

We conclude that the joint bias function of $(\delta, L)$ behaves very similar to the monovariate case that only considers $\delta$: Generally, the multivariate Gaussian bias model describes the data either significantly better or at least equally well to a quadratic bias model. The benefits are most significant for highly biased tracers and large $k_{\mathrm{d}}$.

\section{Bias functions across scales} \label{sec:measureF}

\subsection{The scale dependence of the bias function $f$}
\begin{figure}
    \centering
    \includegraphics[width=\columnwidth]{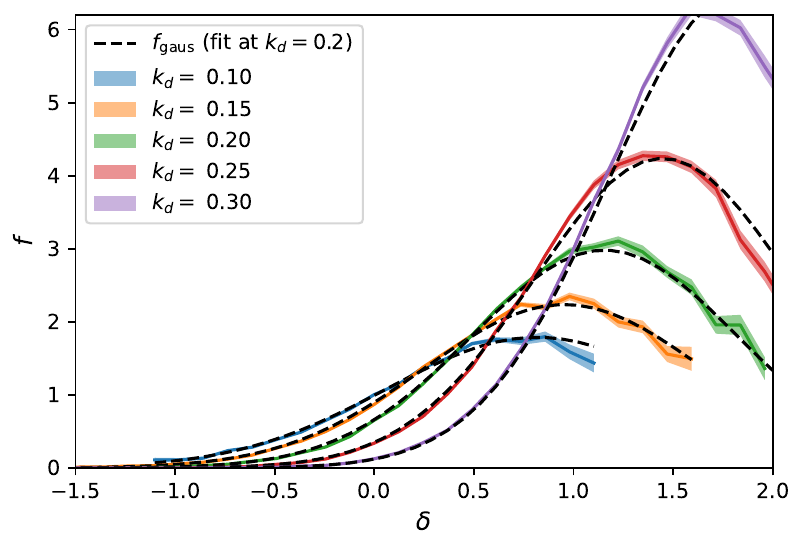}
    \caption{The bias function measured at different smoothing scales in comparison to a multivariate Gaussian bias model fitted at $k_{\mathrm{d}} = 0.2 h \mathrm{Mpc}^{-1}$ and rescaled under the assumption of the PBS. The PBS assumption recovers the bias function at larger scales $k_{\mathrm{d}} \lesssim 0.2 h \mathrm{Mpc}^{-1}$ very well, but seems to be less accurate at smaller scales. }
    \label{fig:fscale_dep}
\end{figure}

\begin{figure*}
    \centering
    \includegraphics[width=0.45\textwidth]{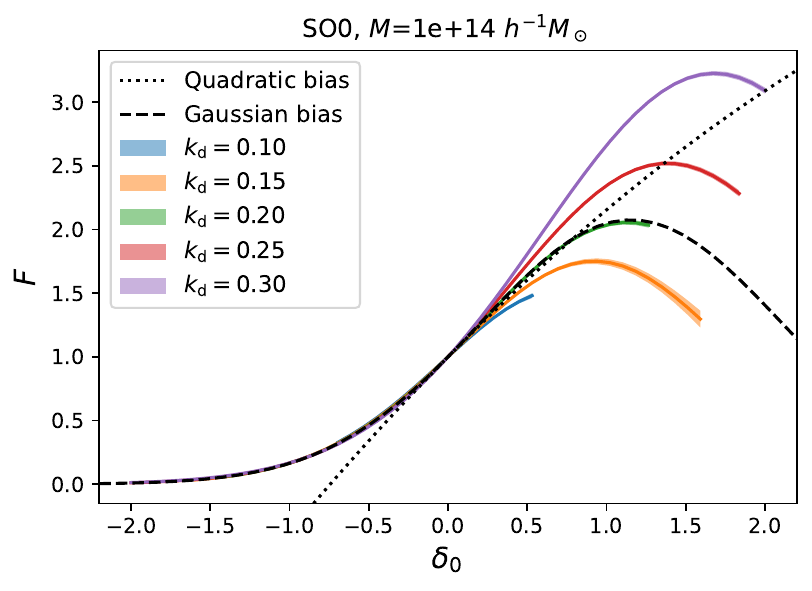}
    \includegraphics[width=0.45\textwidth]{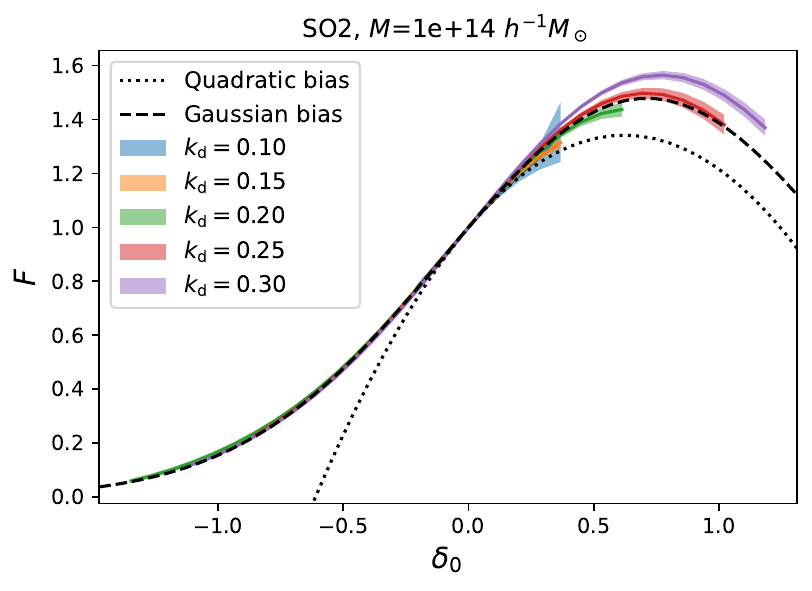}
    \caption{The renormalized bias function of haloes with $M_{200\rm{b}} \sim \SI{e14}{\per \h \msol}$ reconstructed at zeroth spatial order (left) and second spatial order (right) for different damping scales. At second spatial order the reconstructed function is independent of the measurement scale for $k_{\mathrm{d}} \lesssim 0.2 h \mathrm{Mpc}^{-1}$ and it is well described by a Gaussian bias model.}
    \label{fig:biasfunction_scaledep}
\end{figure*}

We have so far focused on measurements of the bias function $f$ at  finite smoothing scales and tested how well different bias models describe $f$ if they are optimized to reproduce the moments of the galaxy environment distribution. However, these tests did not yet ensure that the bias function at smoothing scales larger than the considered scale is well described by the same model. 

Bias functions at different smoothing scales can be unified in a single model through the core assumption of the bias scheme, the Peak-background split (PBS), stating that the response to perturbations is independent of the scale of the perturbation (as long as the scale is sufficiently large). The PBS ensures that if the bias function $f_{\mat{C}_1}$ is defined at some scale with some covariance matrix $\mat{C}_1$ of the perturbations, the bias function at any other scale $f_{\mat{C}_2}$ (with another covariance matrix) follows automatically from $f_{\mat{C}_1}$.

Consider Figure \ref{fig:fscale_dep} where we show the density bias function of haloes with $M_{200\rm{b}} \sim \SI{e14}{\per \h \msol}$ evaluated at different damping scales. In comparison we show the predicted density distribution of a multivariate ($\delta$, $L$) Gaussian bias model with parameters estimated at $k_{\mathrm{d}} = 0.2 h \mathrm{Mpc}^{-1}$, but evaluated with the covariances of the different scales. Note that one can recover the projected density-only version of a multivariate model by projecting the galaxy environment distribution
\begin{align}
    f(\delta) &= \frac{\int p(\delta, L | \mathrm{g}) \mathrm{d}L}{p(\delta)} \nonumber
\end{align}
which leads to a Gaussian with scale-dependent mean and variance depending on the Laplacian bias terms. We notice that the bias functions that are predicted at the scales $k_{\mathrm{d}} \lesssim 0.2 h \mathrm{Mpc}^{-1}$ are very well reconstructed by the PBS assumption. At the scales $k_{\mathrm{d}} = 0.25 h \mathrm{Mpc}^{-1}$ and $k_{\mathrm{d}} = 0.3 h \mathrm{Mpc}^{-1}$ the reconstruction is still reasonably close to the actual function, but clearly not optimal. This means either that the PBS assumption becomes inaccurate at those scales or that additional variables (e.g. $\nabla^4 \delta$) become relevant. Note that this observation is relatively independent on the scale where we have performed the fit (as following considerations show).

Here, we want to understand whether the bias function is well described by a single multivariate Gaussian model across multiple scales. There is a convenient alternative to measuring the bias function at each scale individually and rescaling the bias model accordingly. We may instead measure the aspects of the bias function that should be invariant across the different scales. While this is usually done through the large scale bias parameters, it is also possible to recover this information in a non-parameteric way by directly measuring the renormalized large scale bias function $F$. 

\subsection{Measurements of the renormalized bias function $F$}
\begin{figure}
    \centering
    \includegraphics[width=\columnwidth]{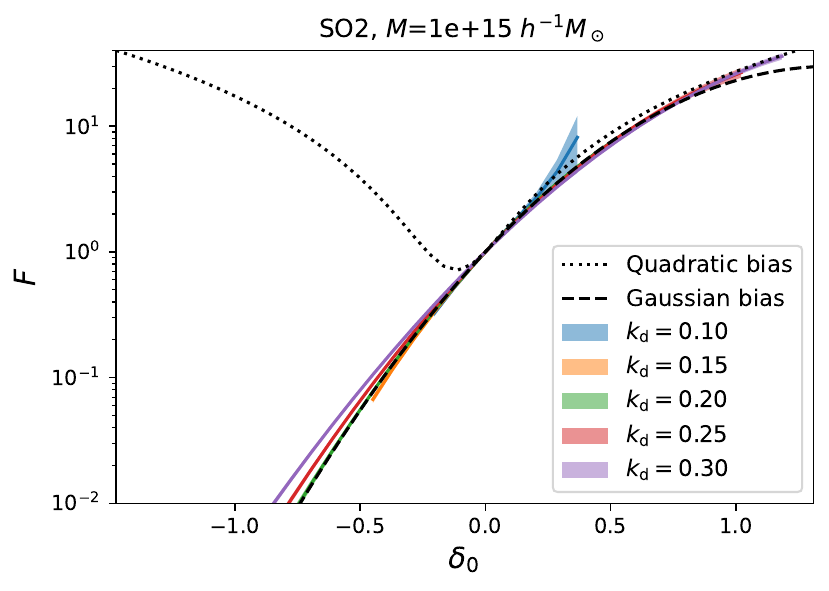}
    \includegraphics[width=\columnwidth]{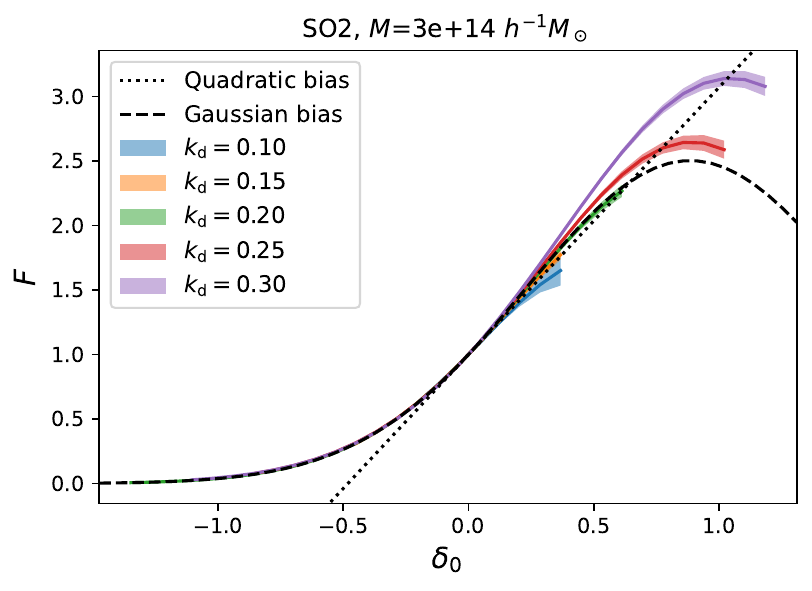}
    \includegraphics[width=\columnwidth]{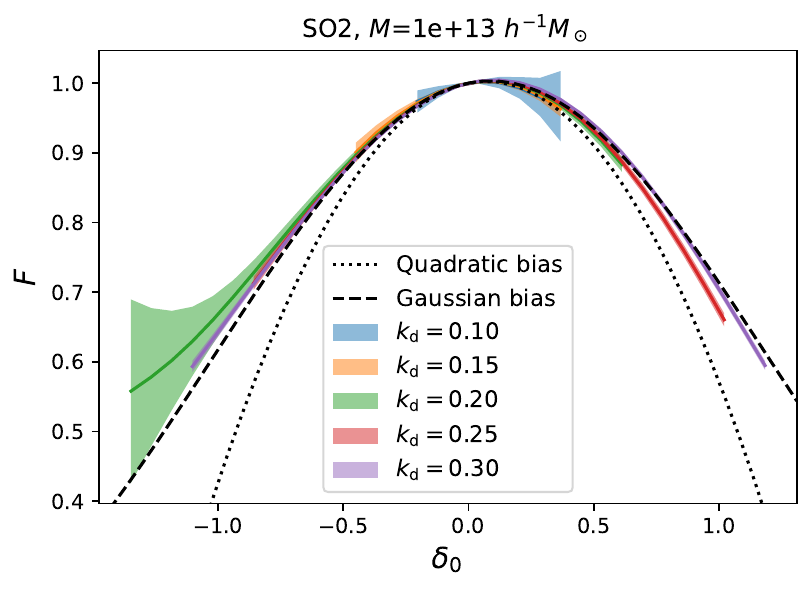}
    \caption{Measurements of the renormalized bias function $F$ for haloes of three different mass selections. A reliable scale-independent reconstruction seems generally possible from damping scales up to $k_{\mathrm{d}} \lesssim 0.2 h \mathrm{Mpc}^{-1}$. The function looks generally better approximated by a Gaussian than a quadratic bias model, especially in regions where the function gets close to zero $F \lesssim 0.5$.}
    \label{fig:Fbiasfunc_vs_masses}
\end{figure}

\begin{figure}
    \centering
    \includegraphics[width=\columnwidth]{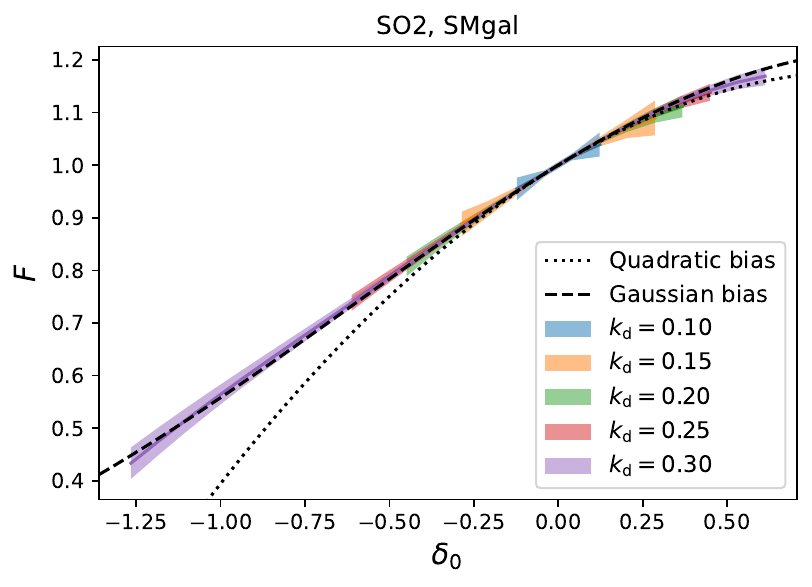}
    \includegraphics[width=\columnwidth]{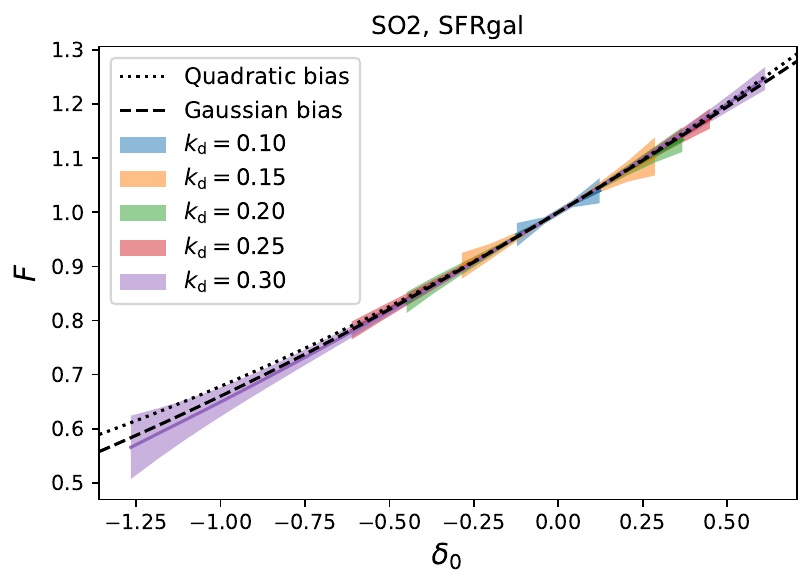}
    \caption{Measurements of the renormalized bias function $F$ for stellar mass selected galaxies (top) and star formation rate selected galaxies (bottom).}
    \label{fig:Fbiasfunc_galaxies}
\end{figure}

As we have shown in Section \ref{sec:measuringF}, we can directly measure the large scale bias function through equation \eqref{eqn:measureF}. We will use this to directly measure $F(\delta_0)$. While the equation should be valid for any value of $\delta_0$, the expectation can be quite dominated by rare outliers with large weights for large values of $\delta_0 / \sigma_0$ . We discuss this in detail for a toy model with known solution in Appendix \ref{app:toymodel}, where we also devise a technique to identify the regime that is unaffected by rare outlier statistics. In plots throughout this section, we will only show this regime that is statistically reliable.

In Figure \ref{fig:biasfunction_scaledep} we show measurements of the renormalized bias function $F(\delta_0)$, comparing the spatial order 0 estimator from equation \eqref{eqn:measureFo0} with the spatial order 2 estimator from equation \eqref{eqn:measureFo2}. The spatial order 0 estimator only reconstructs the function for $\delta_0 \lesssim 0.2$ in a reasonably scale-independent manner, but it clearly yields a very scale-dependent behavior at larger $\delta_0$. This implies that any bias model that is only a function of density (no matter which order) that is fitted at some scale, e.g. $k_{\mathrm{d}} = 0.2 h \mathrm{Mpc}^{-1}$, would not reliably capture the bias at larger scales. Lagrangian local in matter density (LLIMD) models are therefore limited in validity to very large scales, even if they consider polynomials of arbitrary high orders in density.

On the other hand, the spatial order two reconstruction yields an almost perfectly scale-independent behavior at scales $k_{\mathrm{d}} \lesssim 0.2 h \mathrm{Mpc}^{-1}$. Only at smaller scales $k_{\mathrm{d}} = 0.25 h \mathrm{Mpc}^{-1}$ and $k_{\mathrm{d}} = 0.3 h \mathrm{Mpc}^{-1}$ does the reconstruction become scale-dependent, consistent with the observations from Figure \ref{fig:biasfunction_scaledep}. At such small scales the agreement may be improved by considering higher order variables in a reconstruction of fourth spatial order. However, it seems clear that we can reconstruct the renormalized bias function very well on sufficiently large scales with the spatial order 2 estimator.

Where $F$ is reliably reconstructed, it yields a direct estimate of the bias function that can also be measured with separate universe simulations \citep{li_2014, wagner_2015, lazeyras_2016}. Note that the region  $\delta_0 < -1$ is well defined and well behaved, because the $\delta_0$ value corresponds only to a linearly extrapolated density contrast -- describing aspects of the initial conditions, but not the final density. However, we should expect ill-defined behavior for $\delta_0 \gtrsim 1.68$, since a hypothetical separate universe with such a density contrast would collapse to a single point. This point lies anyways far beyond the region that we can reliably reconstruct in a scale-independent manner, so that this limitation is not relevant for the presented measurements.

We compare the measured renormalized bias function with the quadratic and the Gaussian model that is using the bias parameters measured at $k_{\mathrm{d}} = 0.2 h \mathrm{Mpc}^{-1}$. Clearly, both models match the function well at small absolute values $|\delta_0| \lesssim 0.2$. However, while the expansion bias strongly deviates at larger values and catastrophically fails at very small $\delta_0 \lesssim -0.5$, the Gaussian bias seems to be an excellent approximation to the actual function everywhere where it is well measured. 

We measure the renormalized bias function in the same (spatial order 2) manner for three different sets of haloes with masses $\SI{e13}{}$, $\SI{3e14}{}$ and $\SI{e15}{} \SI{}{\per \h \msol}$ in Figure \ref{fig:Fbiasfunc_vs_masses}. The reconstructions up to  $k_{\mathrm{d}} \lesssim 0.2 h \mathrm{Mpc}^{-1}$ seem again reliably scale independent. The Gaussian seems to be a better approximation than the quadratic bias for all three sets of haloes. The difference is most significant in regions where the bias function gets close to zero $F \lesssim 0.5$. 

Finally, we show in Figure \ref{fig:Fbiasfunc_galaxies} the equivalent measurements for galaxies. Similar to our observations from the measurements of the scale-dependent bias functions, the stellar mass selected galaxies are slightly better approximated by a Gaussian, whereas the SFR galaxies are roughly equally well reconstructed by both models. However, it seems that these galaxies have a sufficiently low bias that even a linear bias model would be a good approximation.


We conclude that a multivariate Gaussian bias based on the density and Laplacian model may consistently and accurately describe the bias relation at all scales $k_{\mathrm{d}} \lesssim 0.2 h \mathrm{Mpc}^{-1}$. While we have seen in Section \ref{sec:measure_multivariate} that a Gaussian is still a good approximation to $f$ at scales far beyond $k_{\mathrm{d}} > 0.2 h \mathrm{Mpc}^{-1}$, such scales seem not perfectly reconcilable with the PBS assumption that only assumes $\delta$ and $L$ as variables. It might be possible to push the scale-independence to smaller scales by considering additional variables like $\nabla^4 \delta$, which is however beyond the scope of this paper. 

\section{Why should the bias function be Gaussian?} \label{sec:whygaussian}
\begin{figure}
    \centering
    \includegraphics[width=\columnwidth]{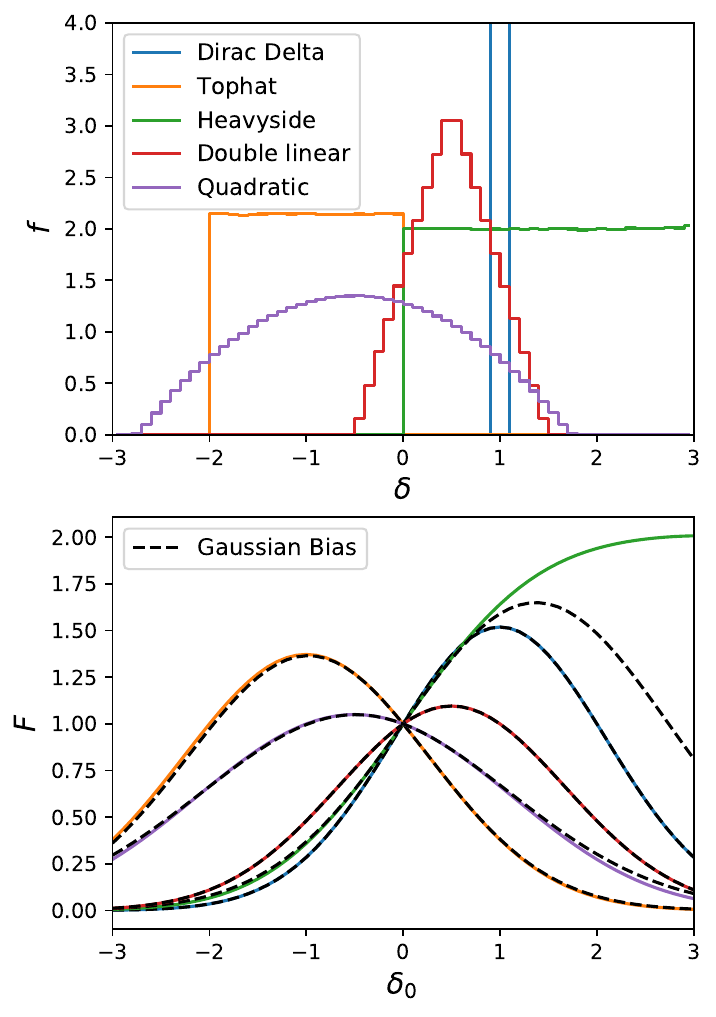}
    \caption{The bias function $f$ (top panel) versus the renormalized large scale bias function $F$ (bottom) for a set of toy models. Selection functions that are very non-Gaussian on small scales naturally lead to bias relations that are close to Gaussian on large scales, since their relation is given by a convolution with a Gaussian.}
    \label{fig:f_vs_F_toy_models}
\end{figure}

We have so far shown through measurements in simulations that the bias function appears close to a Gaussian. A priori it is not obvious why the bias function should appear Gaussian, since it is possible to imagine very complicated galaxy formation scenarios. 

We suggest that the main reason that the large scale bias function appears Gaussian, is that the statistics of the linear density field is itself Gaussian. If the smooth linear density $\delta_1$ on some large length scale is specified, then the conditional distribution of density at any other length scale $p(\delta_2 | \delta_1)$ will follow a Gaussian distribution. Therefore, if we had some non-Gaussian bias function $f_2$ at some small scale, the bias function at larger scales will be $f_2$ convolved with a Gaussian. The resulting bias function $f_1$ will ultimately be closer to Gaussian.

To illustrate this, we consider a set of non-Gaussian selection functions as a toy model. We measure the linear density contrast at a damping scale of $k_{\mathrm{d}} = 0.3 h \mathrm{Mpc}^{-1}$ where $\sigma = 1.1$ and we define different ways to select biased populations based on this. To define these tracers, we use a rejection sampling approach with different probability densities:
{\allowdisplaybreaks 
\begin{align}
    p_{\mathrm{Dirac Delta}} &= \begin{cases}
        1 \,\, & \mathrm{if} \,\,\quad 0.9 < \delta < 1.1 \\
        0 & \mathrm{else}
    \end{cases} \\
    p_{\mathrm{Tophat}} &= \begin{cases}
        1 \,\, & \mathrm{if} \,\,\quad -2 < \delta < 0 \\
        0 & \mathrm{else}
    \end{cases} \\
    p_{\mathrm{Heavyside}} &= \begin{cases}
        1 \,\, & \mathrm{if} \,\,\quad \delta > 0 \\
        0 & \mathrm{else}
    \end{cases} \\
    p_{\mathrm{Double linear}} &= \begin{cases}
        \frac{1}{2} + \delta \,\, & \mathrm{if} \quad -\frac{1}{2} < \delta < \frac{1}{2} \\
        \frac{3}{2} - \delta \,\, & \mathrm{if} \,\,\quad  \frac{1}{2} < \delta < \frac{3}{2} \\
        0 & \mathrm{else}
    \end{cases} \\
    p_{\mathrm{Quadratic}} &= \begin{cases}
        1 - \frac{1}{5} (\delta + \frac{1}{2})^2 \,\, & \mathrm{if} \quad -\frac{1}{2} < \delta < \frac{1}{2} \\
        0 & \mathrm{else}
    \end{cases} \,\,.
\end{align}
}
After sampling the tracers, we measure the bias function $f$ at the corresponding scale as shown in the top panel of Figure \ref{fig:f_vs_F_toy_models}. Further, we measure the renormalized bias function through equation \eqref{eqn:measureFo0} -- representing the bias relation that would be observed at much larger scales. Note that the spatial order 0 reconstruction is perfect in the considered scenario, since the actual selection function depends only on density. Further, recall that $F$ corresponds exactly to $f$ convolved by the Gaussian background distribution at the measured scale:
\begin{align}
    F &= f \circ p \,\,.
\end{align}
We show the renormalized bias function in the bottom panel of Figure \ref{fig:f_vs_F_toy_models} alongside with Gaussian bias models that use the bias parameters measured from the moments of the environment distribution, as in equation \eqref{eqn:b1_o0}. 

While all the small scale selection functions are quite different from Gaussians, the large scale bias function $F$ appears much closer to a Gaussian in almost all scenarios. To understand this, first consider the case of the ``Dirac Delta'' selection function, where $F$ has to be exactly Gaussian. Mathematically, this is the case because a Dirac delta distribution convolved with a Gaussian distribution simply gives a Gaussian distribution centered at the location of the Dirac delta distribution (at $\delta = 1$). Physically, we can interpret $F$ in this situation as the probability to reach exactly the density $\delta = 1$ at $k_{\mathrm{d}} = 0.3 h \mathrm{Mpc}^{-1}$, given that the density contrast is $\delta_0$ at infinitely large scales (normalized to $F=1$ at $\delta_0 = 1$). 

While for none of the other selection functions it is expected that $F$ is \emph{exactly} Gaussian, it is striking that $F$ is quite close to a Gaussian for almost all of them. The convolution simply erases all sharp features of the selection and leaves us with functions that are smooth on the scale $\sigma$ and that are very well approximated by Gaussians. The only case that quite significantly deviates from a Gaussian is the Heavyside selection function. In this scenario, $F$ corresponds exactly to an error-function. However, even in this case, major parts of $F$ ($\delta_0 \lesssim 0.5$) are still very well described.

We may expect that the more compact the selection function is on small scales, the more closely the large scale bias function resembles a Gaussian. From simple toy models -- like e.g. the spherical collapse excursion set formalism \citep{press_1974, bond_1991} we can plausibly understand why the selection function for haloes should be relatively compact. In these models a volume element is considered to be part of a halo of mass $M$ if the density contrast crosses a barrier of $\delta_c = 1.68$ for the first time at the Lagrangian smoothing scale of the halo. The bias function has to appear as a Dirac-delta function at that scale. Deriving the actual bias function on larger scales is slightly more complicated, because it receives some additional scale dependence due to the first-crossing requirement. However, for the case of a sharp $k$ filter the resulting function is
\begin{align}
    F_{\mathrm{ES}} &= \left(1 - \frac{\delta}{\delta_c}\right) \exp \left( \frac{(2 \delta_c - \delta) \delta }{2 \sigma_{M}^2} \right)
\end{align}
as explained e.g. in \citet{Desjacques_2018}, where $\sigma_{M}$ is the variance of the linear density field at the scale of the halo. Given our argument from above, we understand that this is quite close to a Gaussian.

We may further speculate about the reason why the bias function of our mock-galaxies seems to be much wider than that of haloes. Galaxies of some particular selection may occupy haloes of many different masses and appear therefore as a superposition of the bias function of different haloes. It appears that our mocked SFR galaxies occupy a sufficiently wide range in halo masses that the bias function is so smooth that it is hard to tell the difference between e.g. a linear, quadratic or Gaussian bias model.

We conclude that large scale bias functions are likely to be close to Gaussian even if galaxy formation acts in a very non-Gaussian manner on small scales, because small scale and large scale bias are related through a convolution with a Gaussian. More elaborate models beyond the Gaussian paradigm may consider more general basis functions that are optimized to describe the space of positive functions that are smooth at some scale $\sigma$.

\section{Conclusions} \label{sec:conclusions}
In this article, we have presented a new non-parametric method to measure the galaxy environment distribution $p(\delta|\mathrm{g})$, the scale-dependent bias function $f(\delta)$ and its renormalized counter-part $F(\delta_0)$ in Lagrangian space. The new method is very simple and robust, only requiring to evaluate simple expectation values and histograms. We have used the new method to measure the the considered functions and we have found that all three functions are strikingly close to a Gaussian.

Therefore, we have newly introduced a Gaussian Lagrangian bias model. This bias model has the remarkably simple property that all three, $p(\delta|\mathrm{g})$, $f(\delta)$ and $F(\delta_0)$ are given by Gaussian functions with parameters that can easily be expressed in terms of renormalized biases $b_1$ and $b_2$. Further, we have shown that for the multivariate case -- considering e.g. additionally the Laplacian of the density field or the tidal field as variables -- a multivariate Gaussian bias model can be formulated that exhibits the same merits. 

We have systematically tested, how well the scale-dependent bias function is approximated through a Gaussian bias in comparison to a (quadratic) expansion bias model with the same number of free parameters. While the quadratic model generally works only well for some carefully selected regime (e.g. large scales and low-bias objects), the Gaussian seems to be a good approximation in all regimes. In general, the Gaussian appears to be at least as accurate as a quadratic bias (e.g. for low mass haloes and SFR galaxies) or significantly more accurate (e.g. for high mass haloes and small scales), but never notably worse. This is so both for monovariate and multivariate cases. 

Further, we have investigated the scale dependence of the bias function and tested whether a multivariate ($\delta$, $L$) Gaussian bias model that fits the bias function well at one damping scale, also accurately captures the behavior on larger scales. We have found that such consistency between scales can be reliably achieved up to a damping scale of $k_{\mathrm{d}} \sim 0.2 h \mathrm{Mpc}^{-1}$ for any set of tracers. For even larger $k_{\mathrm{d}}$ some additional care must be taken, e.g. by additionally considering higher order derivatives and by carefully testing the reliability of the PBS assumption. 

Beyond the improved accuracy, the Gaussian bias model also has the desirable property that it only predicts positive function values over the whole input domain. This is a physical constraint that is obviously the case for realistic galaxy catalogues, but that is universally violated by the canonical bias expansion approach. This encoded constraint likely contributes significantly to the accuracy of the bias model, especially at small smoothing scales where much of the space may be predicted to be with negative galaxy densities by canonical bias expansion models. Further, it opens up a new set of possibilities which presuppose the positivity of the bias function:
\begin{itemize}
    \item Probability theory can be applied in a rigorous manner with a Gaussian bias model. For example, the galaxy environment distribution forms a well defined probability distribution and the likelihood of observing a set of galaxy environments, given a Gaussian bias model, is a well defined metric and could potentially be used by initial condition reconstruction methods.
    \item With strictly positive predicted galaxy densities, the Gaussian bias could be used to create mock galaxy catalogues with discrete tracers from a given set of bias parameters, by sampling the corresponding probability density. Predictions of discrete tracers would mimic actual galaxy catalogues with a notably larger degree of realism than the continuous fields that are typically predicted by bias models. However, it would be important to develop a correlated way of sampling that allows to create super- or sub-Poissonian statistics.
\end{itemize}
For the latter point, we note that such an option also exists when using non-parametric measurements of the bias function that we have presented in this article. For example, one could measure the bias function of galaxies from some small volume hydro simulation and use it to populate a large scale dark matter only simulation.

While the Gaussian bias model seems to be a promising alternative to polynomial expressions, there still remain a few open questions that may be addressed in future studies. An important aspect of bias models to describe most measurable statistics (like e.g. the power spectrum) is a description of the large scale contribution of small scale stochastic components. Conveniently, the galaxy environment distribution, that we have investigated here, does not include any unaccounted stochastic component from small scales, 
so that we could investigate the deterministic part of the bias relation without modelling stochastic terms. For other statistics, the stochastic components may probably be modelled for a Gaussian model in the same way they are typically modelled for expansion biases, but this aspect deserves some additional theoretical consideration. In particular, it would be an interesting alternative option to try mimicking the stochastic components through a discrete sampling process as mentioned above.

Further, it would be interesting to consider higher order generalizations that naturally recover a Gaussian bias at second order. In \citetprob{} we have introduced the cumulant bias expansion as a general description of the large scale bias function $F$ and suggested a few practical suggestions to implement it into actual bias functions. However, it is difficult to find a general method to describe probability distributions with more than two cumulants in a way that ensures the positivity $f > 0$. Therefore, the question of what is an optimal basis at orders $n > 2$ remains open and can hopefully be addressed by future studies.

We conclude that a Gaussian bias model provides a mathematically simple, well behaved and accurate description of the bias function. If computationally feasible, it should generally be preferred over a second order expansion. For cases where practical aspects may favor an expansion approach\footnote{For example for a bias expansion, the power spectrum can be decomposed into a finite set of cross spectra which only need to be calculated once, but not for every set of parameters.}, this knowledge can still be used by re-expressing canonical higher order parameters through the first two cumulant biases, as explained in \citetprob.

\begin{acknowledgements}
The authors thank Sergio Contreras and Sara Ortega Martinez for providing mocks of galaxies.  The authors thank Oliver Philcox for helpful discussions and comments to the draft and Simon White and Oliver Hahn for helpful discussions. We acknowledge funding from the Spanish Ministry of Science and Innovation through grant number PID2021-128338NB-I00. 
RV acknowledges the support of the Juan de la Cierva fellowship (FJC2021-048002-I). MPI is supported by STFC consolidated grant no. RA5496.
\end{acknowledgements}

\bibliographystyle{aa} 
\bibliography{gaussian_bias} 

\newpage

\begin{appendix} 
\section{Measuring the bias function for a toy model} \label{app:toymodel}

\begin{figure}
    \centering
    \includegraphics[width=\columnwidth]{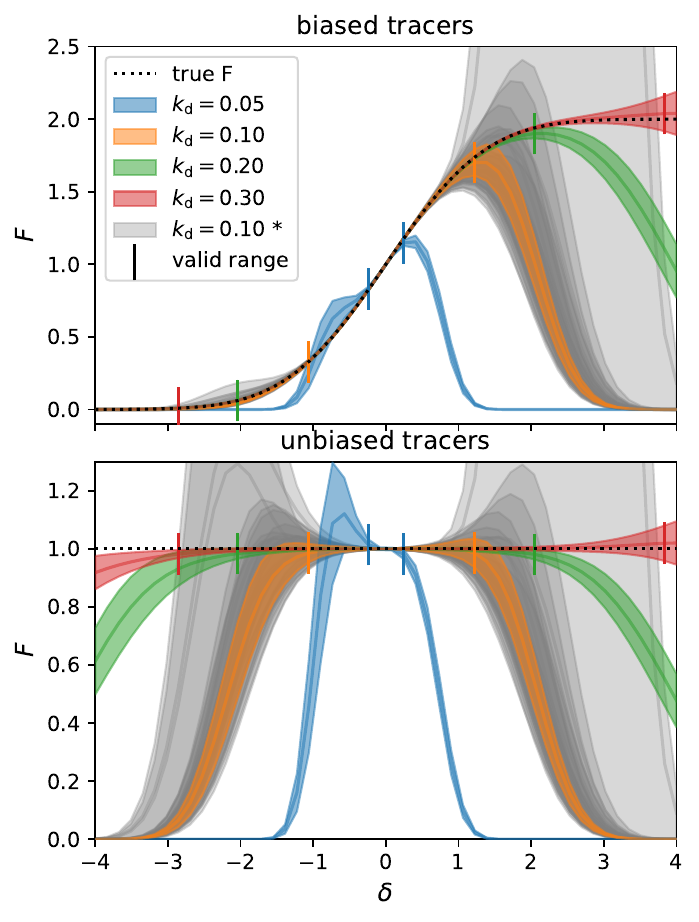}
    \caption{The large scale bias function $F$ measured for a toy model with known solution (top) and for unbiased tracers (bottom). Coloured shaded regions show the measurement (and jackknife error) at different damping scales and the grey regions show different realizations at $k_{\mathrm{d}} = 0.1 h \mathrm{Mpc}^{-1}$. At larger damping scales the measurement gets generally biased at smaller values of $\delta$. The jackknife errors underestimate the uncertainty, since the measurement is sensitive to rare outliers, but we can estimate the validity range of the measurements as where unbiased tracers have $F \sim 1$.}
    \label{fig:toy_biasfunc}
\end{figure}

Here, we consider a toy model to establish the range of validity of our measurements of the large scale bias function $F$. As explained in Section \ref{sec:measuringF}, the estimator of $F$ is given by 
\begin{align}
    F(\delta_0) &= \exp \left(\frac{-\delta_0^2}{2 \sigma^2} \right) \expectgal{\exp \left(\frac{\delta \delta_0}{\sigma^2} \right)} \label{eqn:app_measureF}
\end{align}
for the monovariate density-only scenario. This expectation value is very sensitive to rare outliers. For example, at $\delta_0 = 1$ and $\sigma=0.5$ a $5 \sigma$ outlier galaxy with $\delta = 2.5$ would contribute with a weight that is $\exp(5) \sim 150$ times higher than the weight of a galaxy with $\delta = 0$. What makes matters worse is that a simulation does not perfectly follow Gaussian statistics. The density distribution at scales close to the fundamental frequency of the box is quite non-Gaussian and it is relatively unlikely to exhibit a realistic number of outliers. 

To investigate this effect we set up an example set of biased tracers. For this we select all particles that have at a small damping scale $k_{\mathrm{d}} = 0.3 h \mathrm{Mpc}^{-1}$ a density $\delta$ higher than 0, corresponding to exactly half of the tracers. Therefore, $f = 2 \Theta(\delta)$ is given at this scale by a heavy-side function $\Theta$ and the large scale bias function is then given by
\begin{align}
    F(\delta_0) &= \int_{0}^{\infty} p(\delta - \delta_0) f(\delta) \mathrm{d} \delta \nonumber \\
                &= 1 + \mathrm{erf} \left( \frac{\delta_0}{\sqrt{2} \sigma} \right)
\end{align}
where the factor $2$ comes from the normalization constraint $F(0) = 1$ and where $\sigma^2$ is the variance at $k_{\mathrm{d}} = 0.3 h \mathrm{Mpc}^{-1}$. Since the set of biased tracers depends exactly only on density, we should be able to recover them exactly by applying equation \eqref{eqn:app_measureF}  at any damping scale (keeping the set of tracers fixed). We show such measurements for different scales in the top panel of Figure \ref{fig:toy_biasfunc}. The measurements at different scales agree excellently at small $\delta_0$. However, the smaller $k_{\mathrm{d}}$, the earlier in $|\delta_0|$ the measurements tend to deviate from the correct function. Noteworthy, the deviations are much larger than the jackknife error estimates. However, given the extreme sensitivity to outliers and the effectively low number of truely independent samples (due to the limited number of modes at large scales), it is very likely that the Jackknife sampling never sees the rare outliers that carry the expectation value, and therefore strongly underestimates the statistical error. To highlight this, we draw as grey contours the uncertainty ranges of 12 different realizations of this setup evaluated at $k_{\mathrm{d}} = 0.1 h \mathrm{Mpc}^{-1}$. Most of these go down at a similar value of $\delta$ and also with a small error estimate. However, there is one realization which gets extremely large, with a maximum of $F \sim 26$ at $\delta_0 = 2.7$ which is far outside the range of the plot. The average of a large number of realizations will still give the correct value at $\delta_0 = 2.7$. However, the distribution over realizations is so imbalanced that it is very difficult to access the true statistical uncertainty from only a single realization.

Therefore, to estimate the range of validity of our bias function measurements, we access the values of $F$ that we would determine for an unbiased set of tracers. For this we average equation \eqref{eqn:app_measureF} over the full Lagrangian grid. For the pure density case the result is shown in the bottom panel of Figure \ref{fig:toy_biasfunc}. If the distribution on the grid was a perfect Gaussian, it would be $F=1$, but due to the aforementioned non-Gaussianity and the sensitivity to outliers, the measurements deviate from the correct $F$ in a similar way to the top panel. Therefore, we define as the range of validity, the range where $|F - 1| < \epsilon$ when $F$ is averaged over the whole grid and where we choose $\epsilon = 0.02$. We mark the corresponding points in Figure \ref{fig:toy_biasfunc} and we find that $F$ is measured quite reliably for this selection across all damping scales. For all figures in the main-text we only show measurements in this reliable regime.

While we use only simulations with fixed amplitudes of modes as in \citet{angulo_pontzen_2016}, we have repeated the experiment in this section for both fixed and unfixed initial conditions and we found no significant difference in the range of validity. 

\section{Derivation of the renormalized Gaussian tidal bias} \label{app:renormtidal}
We want to find the renormalized form for a multivariate Gaussian bias function $f$ that uses $\delta$, $\mat{K}$ and $L$ as variables. We make the Ansatz for its large scale limit
\begin{align}
    \log F(\delta_0, L_0, \mat{K}_0) 
           &= \log F_0(\delta_0, L_0) + \log F_{\mat{K}}(\mat{K}_0) \\
    \log F_{\mat{K}}(\mat{K}_0) &= b_{K^2} K_0^2 \label{eqn:app_Fk}
\end{align}
where $F_0$ is the function that we have already derived in ... and where $K^2 = \trace{\mat{K} \mat{K}}$. Note that this is the only possible Ansatz that is at most second order in the considered variables and obeys the isotropy constraint. Since both $F$ and the background distribution $p(\delta, L, K) = p(\delta, L) p(K)$ factorize, also the scale dependent bias function factorizes in the same manner
\begin{align}
    f(\delta, L, \mat{K}) &= f_0(\delta, L) f_{\mat{K}}(\mat{K})
\end{align}
and it is sufficient to enforce the renormalization constraint in $(\delta, L)$ and in $\mat{K}$ individually. Therefore, we only have to find a function $f_{\mat{K}}$ such that
\begin{align}
    F_{\mat{K}}(\mat{K}_0) &= \expect{f_{\mat{K}}(\mat{K} + \mat{K}_0)} \,\,.
\end{align}
We start with the assumption that $f_{\mat{K}}$ corresponds to a multivariate Gaussian with a flexible normalization $A$:
\begin{align}
    f_{\mat{K}}(\mat{K}) &= A \cdot N(\mat{K}, \mat{\mu}=0, \mat{C}_f)
\end{align}
where $\mat{C}_f$ is a rank four tensor and the mean $\mat{\mu}$ has to be zero, because of the isotropy + tracelessness constraint and we are to determine $A$ and $\mat{C}_f$. The pdf of a zero-mean multivariate normal distribution over a symmetric traceless tensor $\mat{K}$ (with five degrees of freedom) can be written as
\begin{align}
    N(\mat{K}, \mat{\mu}=0, \mat{C}) &= \frac{1}{\sqrt{(2\pi)^5 |\det \mat{C}|}} \exp \left( -\frac{1}{2} \mat{K}^T \mat{C}^+ \mat{K} \right)
\end{align}
where $\mat{C}^+$ is the pseudo inverse of $\mat{C}$ and \quotes{$\det$} indicates a pseudo-determinant -- that is the product of the non-zero eigenvalues of $\mat{C}$. Formally this pdf is not a distribution over all 9 components of $\mat{K}$, but only describes the variation of the probability density in the five dimensional subspace where $\mat{K}$ is symmetric and traceless. The background distribution also corresponds to a multivariate Gaussian
\begin{align}
    p(\mat{K}) &= N(\mat{K}, 0, \mat{C}_{\mat{K}}) \\
    \mat{C}_{\mat{K}} &= \frac{2 \sigma^2}{15} \mat{J}_{2=2}
\end{align}
where $\mat{J}_{2=2}$ is the isotropic rank four tensor with indices
\begin{align}
    \mat{J}_{2=2, ijkl}  &= \frac{1}{2} (\delta_{ik} \delta_{jl} + \delta_{il} \delta_{jk}) - \frac{1}{3} \delta_{ij} \delta_{kl}
\end{align}
where here $\delta_{ij}$ is the Kronecker delta symbol. The symmetry requirements that arise to this form are explained in more detail in Section 4 of \citetprob. Therefore, we have
\begin{align}
    F_{\mat{K}} &=  \int p(\mat{K}) f_{\mat{K}}(\mat{K} + \mat{K}_0) \mathrm{d}^5 \mat{K} \nonumber \\
        &=  \int p(\mat{K}_0 - \mat{K}) f_{\mat{K}}(\mat{K}) \mathrm{d}^5 \mat{K}
\end{align}
where we have used a substitution plus the symmetry of $p$ to show that this corresponds to a convolution
between $p$ and $f_{\mat{K}}$. The convolution of two multivariate Gaussian leads to a new multivariate Gaussian with their mean and their covariances added together:
\begin{align}
    F_{\mat{K}}(\mat{K}_0) &= A N(\mat{K}_0, 0, \mat{C}_F = \mat{C}_{f} + \mat{C}_{\mat{K}}) \\
      &= \frac{A}{\sqrt{(2\pi)^5 \left| \det(\mat{C}_{F}) \right|}} \exp \left(- \frac{1}{2} \mat{K}_0^T \mat{C}_F^{+} \mat{K}_0 \right) \,\,.
\end{align}
Identifying terms with equation \eqref{eqn:app_Fk} leads to
\begin{align}
    \mat{C}_{\mat{F}}^{+} &= - 2 b_{K^2} \mat{J}_{2=2} \\
    A &= \sqrt{\left| (2 \pi)^5 \det(\mat{C}_{F}) \right|} \\
    \mat{C}_{\mat{b}}  &= \mat{C}_{\mat{F}} - \mat{C}_{\mat{K}} \nonumber \\
    &=  - \frac{1}{2 b_{K^2}} \mat{J}_{2=2} - \frac{2 \sigma^2}{15} \mat{J}_{2=2} \nonumber \\
    &= - \frac{1 + \frac{4\sigma^2}{15} b_{K^2}}{2 b_{K^2}}  \mat{J}_{2=2}
\end{align}
where we have used that $\mat{J}_{2=2}$ is its own pseudo-inverse. We find
\begin{align}
    f_{\mat{K}} &= \sqrt{\left| \frac{\det \mat{C}_F}{ \det \mat{C}_{\mat{b}} } \right|} \exp \left(- \frac{1}{2} \mat{K}^T \mat{C}_b^{+} \mat{K} \right) \nonumber \\ 
      &= \left|1 + \frac{4 \sigma^2}{15} b_{K^2} \right|^{-5/2} \exp \left(\frac{b_{K^2} K^2}{1 + \frac{4}{15} \sigma^2 b_{K^2}} \right)
\end{align}
where we have used that $\mat{J}_{2=2}$ has five non-zero eigenvalues that are equal to 1 so that the pseudo determinant is
\begin{align}
    \det (a \mat{J}_{2=2}) = a^5.
\end{align}

\end{appendix}


\end{document}